\documentclass[]{spie}  %>>> use for US letter paper
\usepackage{appendix}
\usepackage[]{graphicx}
\usepackage{float}
\usepackage{multirow}
\usepackage{mathtools}
\graphicspath{ {./figures/} }

\newcommand{\tabcell}[1]{\begin{tabular}{@{}c@{}}#1\end{tabular}}

\newcommand{\arcsec}{''}

\title{The InfraRed Imaging Spectrograph (IRIS) for TMT: photometric precision and ghost analysis} 

\author{
Nils Rundquist\supit{a,b},
Gregory Walth\supit{a}, 
Shelley A. Wright\supit{a,b}, 
Ryuji Suzuki\supit{g}, 
Tuan Do\supit{c},  
Edward L. Chapin\supit{e}, 
Eric Chisholm\supit{f}, 
Jennifer Dunn\supit{e}, 
Yutaka Hayano\supit{g} , 
Chris Johnson\supit{c}, 
James E. Larkin\supit{c}, 
Reed L. Riddle\supit{d}, 
Matthias Schoeck\supit{e,f}, 
Ji Man Sohn\supit{c}
\skiplinehalf
\supit{a} Center for Astrophysics \& Space Sciences, University of California San Diego, USA; \\
\supit{b} Department of Physics, University of California San Diego, USA; \\
\supit{c} Physics \& Astronomy Department, University of California Los Angeles, CA 90095 USA; \\
\supit{d} Caltech Optical Observatories, 1200 E California Blvd., Pasadena, CA 91125 USA; \\
\supit{e} National Research Council of Canada - Herzberg, Victoria, BC, V9E 2E7 Canada; \\
\supit{f} Thirty Meter Telescope Observatory Corporation, Pasadena, CA 91105 USA; \\
\supit{g} National Astronomical Observatory of Japan, Osawa, Mitaka, Tokyo, 181-8588 Japan \\
}

\authorinfo{Further author information: (Send correspondence to N.R.)\\N.R...: E-mail: nrundqui@ucsd.edu}
\begin{document} 
  \maketitle 

%%%%%%%%%%%%%%%%%%%%%%%%%%%%%%%%%%%%%%%%%%%%%%%%%%%%%%%%%%%%% 
\begin{abstract}

The InfraRed Imaging Spectrograph (IRIS) is a first-light instrument for the Thirty Meter Telescope (TMT) that will be used to sample the corrected adaptive optics field by NFIRAOS with a near-infrared (0.8 - 2.4 $\mu$m) imaging camera and Integral Field Spectrograph (IFS).  In order to understand the science case specifications of the IRIS instrument, we use the IRIS data simulator to characterize photometric precision and accuracy of the IRIS imager.  We present the results of investigation into the effects of potential ghosting in the IRIS optical design.  Each source in the IRIS imager field of view results in ghost images on the detector from IRIS's wedge filters, entrance window, and Atmospheric Dispersion Corrector (ADC) prism.  We incorporated each of these ghosts into the IRIS simulator by simulating an appropriate magnitude point source at a specified pixel distance, and for the case of the extended ghosts redistributing flux evenly over the area specified by IRIS's optical design. We simulate the ghosting impact on the photometric capabilities, and found that ghosts generally contribute negligible effects on the flux counts for point sources except for extreme cases where ghosts coalign with a star of $\Delta$m$>$2 fainter than the ghost source.  Lastly, we explore the photometric precision and accuracy for single sources and crowded field photometry on the IRIS imager.

\end{abstract}

%>>>> Include a list of keywords after the abstract 


%%%%%%%%%%%%%%%%%%%%%%%%%%%%%%%%%%%%%%%%%%%%%%%%%%%%%%%%%%%%%
\section{INTRODUCTION}\label{sec:intro}  % \label{} allows 

With three giant segmented-mirror telescopes (GSMTs) on the horizon new scientific opportunities will be available to astronomers by utilizing unprecedented spatial resolution and photometric capabilities. The Thirty Meter Telescope (TMT) instrumentation aims to sample the diffraction-limit of a 30-m aperture with high precision relative photometric accuracy. The InfraRed Imaging Spectrograph (IRIS)\cite{Larkin1,Larkin2,Larkin3,Larkin4} is a first-light instrument for TMT which will enable the use of TMT for near-infrared (0.8 - 2.4 $\mu$m) imaging and integral-field spectroscopy (IFS). The imager operates with a plate scale of 0.004\arcsec per pixel with a maximum field of view of 34\arcsec x 34\arcsec \cite{imager1,imager2}. IRIS imager and integral field spectrograph will utilize a real-time data reduction pipeline\cite{Walth2} for data processing. Characterizing and predicting IRIS's photometric precision and accuracy is imperative for aiding the design of the instrument, adaptive optics system, and the data reduction system\cite{Walth1}.

It is important to understand potential "ghost images" (i.e., additional point or resolved images generated by a multi-layer optical system) introduced by the optical design of IRIS that can directly impact the astrometric and photometric accurary. The photometric budget is defined by the combination of telescope optics, Narrow Field InfraRed Adaptive Optics System (NFIRAOS)\cite{adc1} \cite{adc2}, and IRIS optical system. Each of these optical systems have the potential capability to redistribute flux on the detector field of view (FoV), causing ghost images from science sources. The observed source magnitude and its optical location determines the brightness and location of the ghosts that are observed by the detector. It is critical to understand the different ghost impacts to properly characterize scenarios that may cause ghost images to adversely affect IRIS science cases.  The existence of ghost images also increases the overall noise seen within the detector FoV.  Ghost images therefore have a direct impact on the total photometric precision achievable by IRIS, and increases the base noise-floor of the imager and IFS. It is therefore necessary to  analyze the overall impact of ghosts on photometric precision and noise-floor contribution in order to accurately define IRIS's photometry budget.

We use the IRIS data simulator\cite{Wright1,Wright2,Wright3} to analyze the photometric effects of ghost images on the IRIS imager. Using the current specifications for IRIS's optical design and filter characteristics, we simulate the effects of the optical system, IRIS throughputs, sky background, ghosts, and Poisson noise. We simulated point sources of magnitudes 1-25 (Vega), with 25 iterations per simulation, including and not including the effects of IRIS in order to assess the photometric error for each simulated source using aperture photometry and Point Spread Function (PSF) fitting routine \textit{Starfinder}\cite{starfinder}. \textit{Starfinder} returns photometric results by extracting a PSF from the image, and fitting that PSF to each source in the field to estimate flux values for each source. This paper reports on the results of the simulations done, analyzing the specific ghost image photometry for both isolated ghost images and point source, binary, and crowded field science cases with the IRIS imager.

\section{Ghost Image Simulations}

We have developed an extensive data simulator for the IRIS imager which includes the effects of TMT, NFIRAOS, and IRIS optical systems\cite{Do2014,Wright3}. We have included additional effects of ghost images, and simulate 25 iterations of single point source and multiple point-source cases with magnitudes 1-25 in K-band (2 - 2.4 $\mu$m) wavelengths. For error analysis, we also simulate ideal models that do not include noise and ghost images in the simulation. We infer the photometric error introduced by ghost images by comparing the flux of the simulated IRIS sources with the known theoretical flux of a given magnitude source using both aperture and \textit{Starfinder} photometry, given by
\begin{equation}
    \mathrm{Photometric\ Error\ } (\%) = 100* \frac{F_M - F_{*}}{F_M}
    \label{eq1}
\end{equation}
where $F_M$ is the known flux from the model image, and $F_{*}$ is the IRIS image flux.

Through aperture photometry we analyze the photometric effects of ghost images as a function of radius from a simulated point source.  We also use \textit{Starfinder} to calculate the absolute photometric error of IRIS for high signal-to-noise ratio (SNR). We simulate each isolated ghost image to assess the total photometric error associated with each ghost type as well as varying science cases to estimate the impact of ghosts in IRIS.

\subsection{Types of Ghost Images}

We investigate three potential types of ghost images generated by the optical system: the TMT Entrance Window, the Atmospheric Dispersion Correction (ADC) prism, and IRIS's wedge-shaped filters. We use \textit{Starfinder} and aperture photometry on each simulated ghost to estimate its photometric impact.

\subsubsection{Entrance Window Ghost Images}
The TMT Entrance Window generates even flux distribution ghosts located directly over the science source observed in the field. The ghost flux is a top-hat distribution 0.8 arcseconds in diameter, with surface brightness of 9.4 magnitudes per square arcsecond (Vega) greater than the field source. Figure 1 illustrates the effect of the entrance window ghost.
 
 \begin{figure}[h!]
    
     \centering
     \includegraphics[scale=0.7]{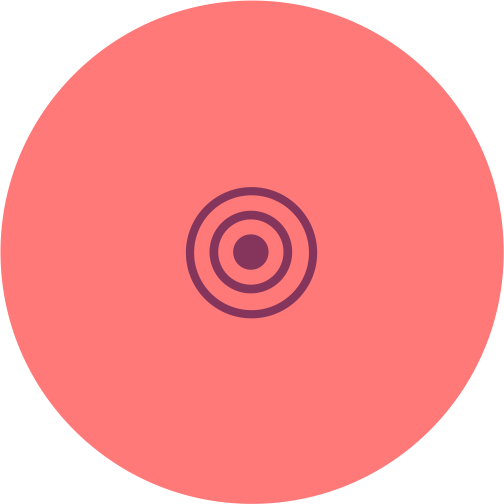}
     \includegraphics[scale=0.6]{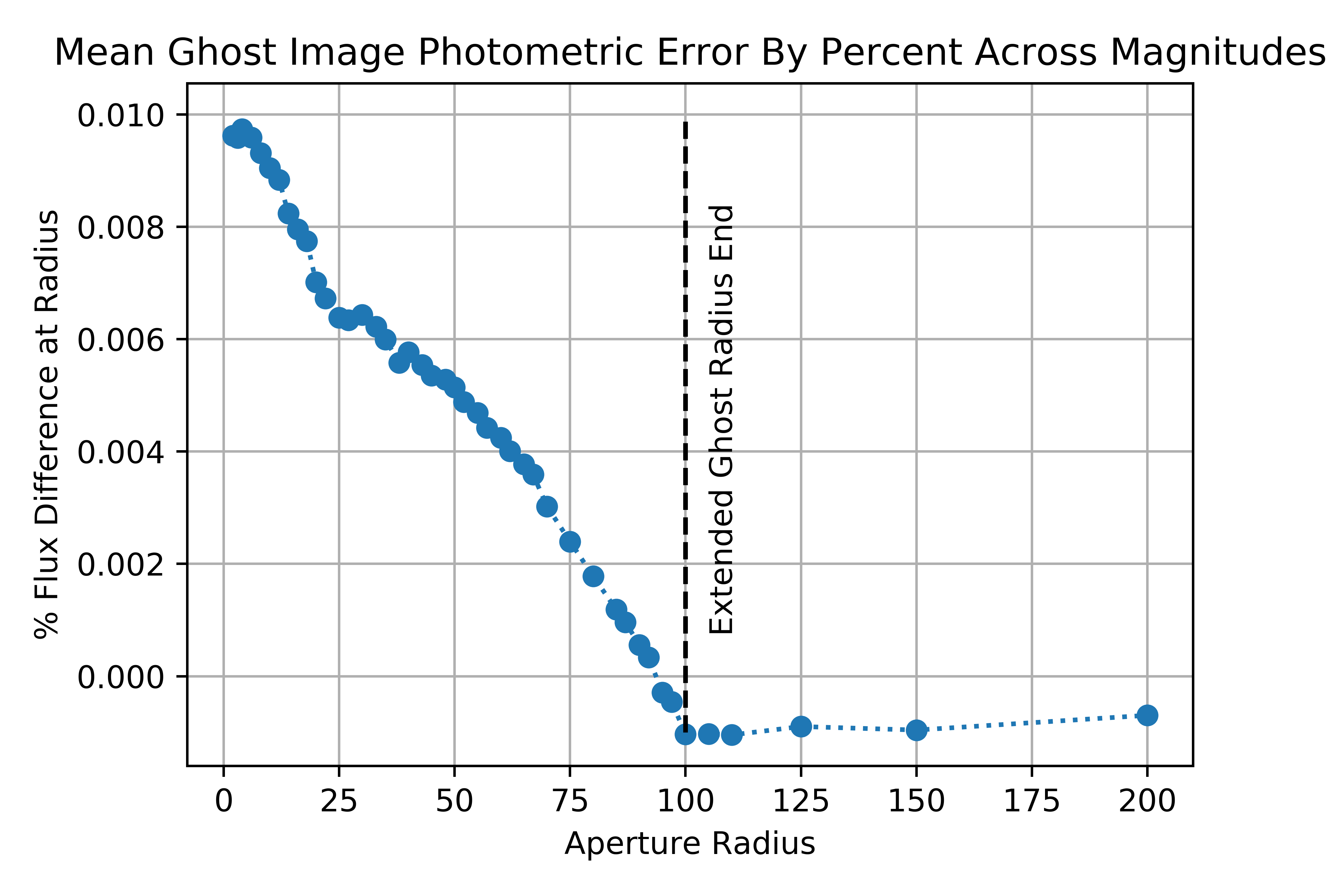}
     \caption{\textbf{Left:} Representation of the Entrance Window Ghost Image. The dark circles represent the Airy pattern of a point source on the IRIS imager.  The red area represents the extent of the Entrance Window ghost image. \textbf{Right:} The photometric error associated with the Entrance Window Ghost in isolation as a function of radius, averaged across each magnitude point source simulation.  Photometric error approaches 0 as the aperture encompasses the affected area. Each photometric error point is the result of 625 IRIS point source simulations, in addition to those simulated for point source comparison.}
     \label{fig:entranceghost}
 \end{figure}

The Entrance Window ghost is the dimmest of the predicted ghost images. With surface brightness 9.4 magnitudes ($\Delta$m=+9.4) per square arcsecond fainter than its source, affecting a 0.8 arcsecond-diameter circle (0.502 square arcseconds), the ghost has a cumulative photometric impact of roughly half of a point source of +9.4 magnitudes. At the 4 miliarcsecond (mas) per pixel spatial scale of the IRIS imager, the Entrance Window ghost corresponds to a 100-pixel-radius circle centered over the originating point source, with flux evenly distributed over this area, which results in small flux counts distributed per pixel. Using \textit{Starfinder} we estimate a total photometric error of 0.010$\pm$0.001\% from the Entrance Window.

\subsubsection{ADC Prism Ghost Images}
The ADC generates ghost images that are top-hat flux distributions that are 0.18 arcseconds in diameter (0.025 square arcseconds). ADC ghosts are located at a distance of 23 arcseconds from the originating source, corresponding to 4750 pixels using the IRIS imager's 4 mas/pixel scale, with a surface brightness of 5.5 magnitudes per square arcsecond greater than the field source.  Figure 2 illustrates the effect of the ADC ghost.

\begin{figure}[h!]
    
    \centering
    \includegraphics[scale=0.4]{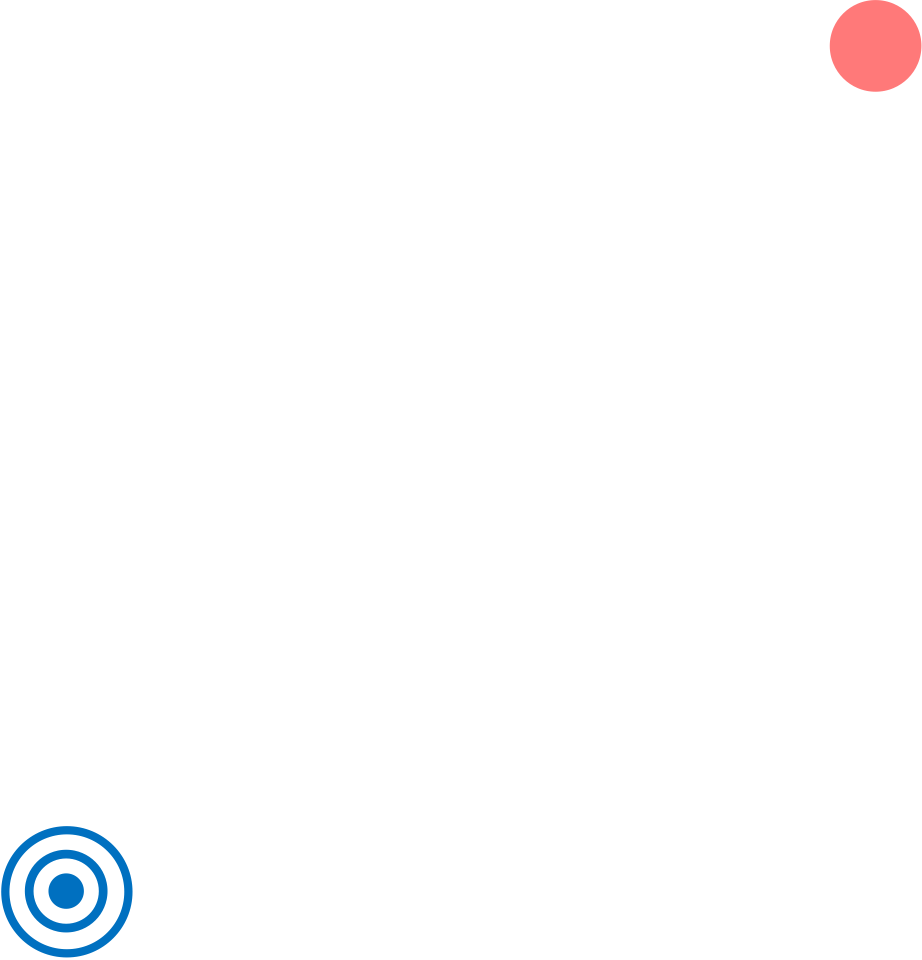}
    \includegraphics[scale=0.6]{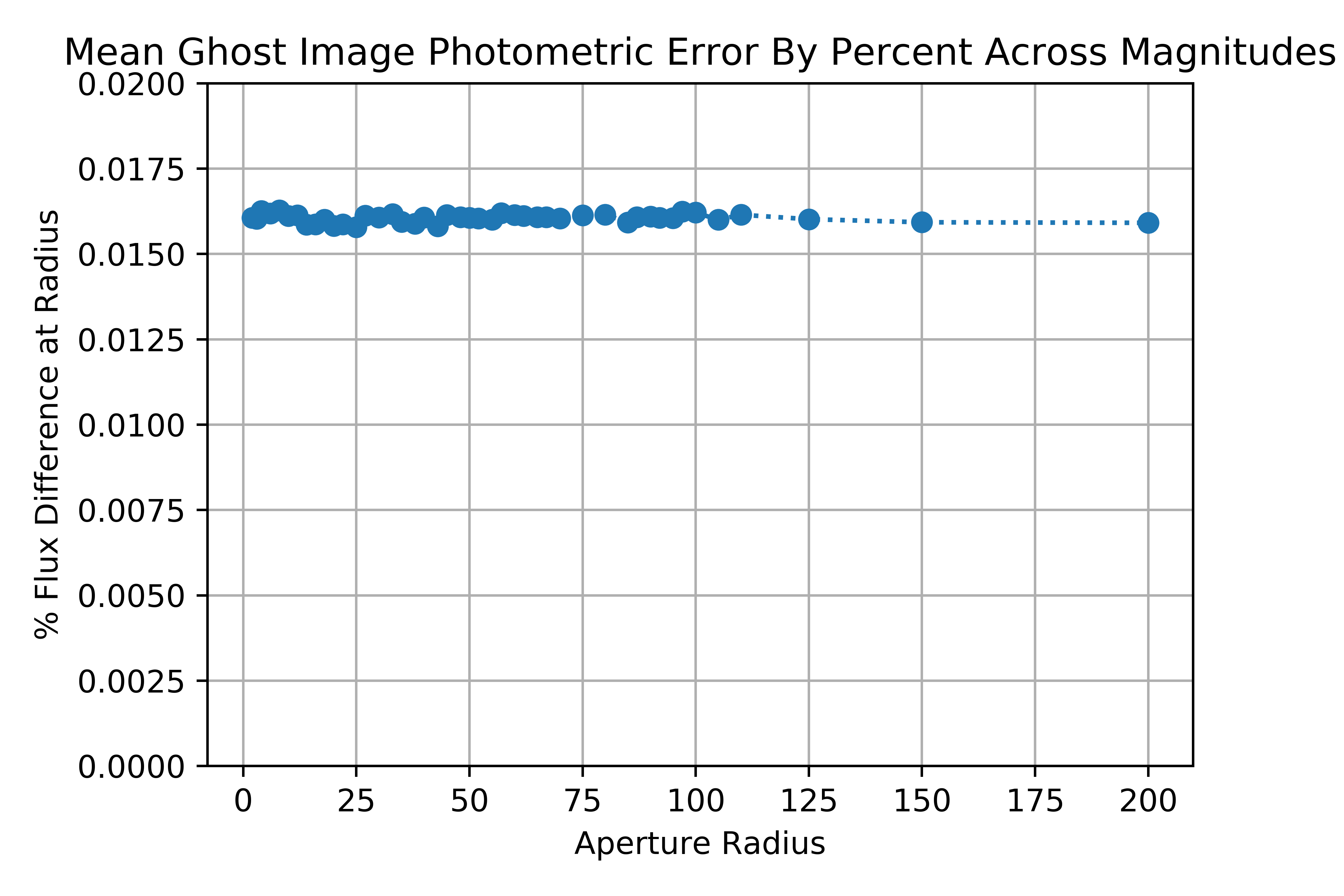}
    \caption{\textbf{Left:} The ADC ghost image is represented by the red area relative to the blue originating point source, at a distance of 23 arcseconds (4750 pixels at 4 mas/pixel). \textbf{Right:} The photometric error associated with the ADC ghost in isolation as a function of radius. Photometric error remains constant as aperture increases due to the ghost image's great distance from the source.}
    \label{fig:adcghost}
\end{figure}

The ADC ghost image is the brightest ghost image in terms of magnitude difference, contributing $\sim$2.5\% the flux of a point source of $\Delta$m=+5.5 magnitudes fainter than the originating source. Using \textit{Starfinder}, we estimate the total photometric impact of this ghost at 0.0173$\pm$0.005\%. The position angle of the ADC ghost relative to its source object changes as the ADC prism rotates, although distance remains constant. Due to the large distance of the ADC ghost relative to its originating source, it is important to consider the potential for this ghost to be projected over a science object. We explore this possibility and the associated photometric impact for science objects of varying magnitude difference in section 2.2.1.

\subsubsection{Wedge Ghost Images}
Double reflections from IRIS's wedge-shaped filters results in a point-source-like ghost image seen 63 mas from the source in the field, assuming the 10 arcsecond filter-wedge angle.  This corresponds to 15.75 pixels at the IRIS imager's 4 mas/pixel scale. Figure 3 illustrates the wedge point-source ghost. 

\begin{figure}[h!]
    \centering
    
    \includegraphics[scale=1.3]{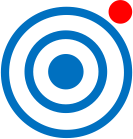}
    \includegraphics[scale=0.5]{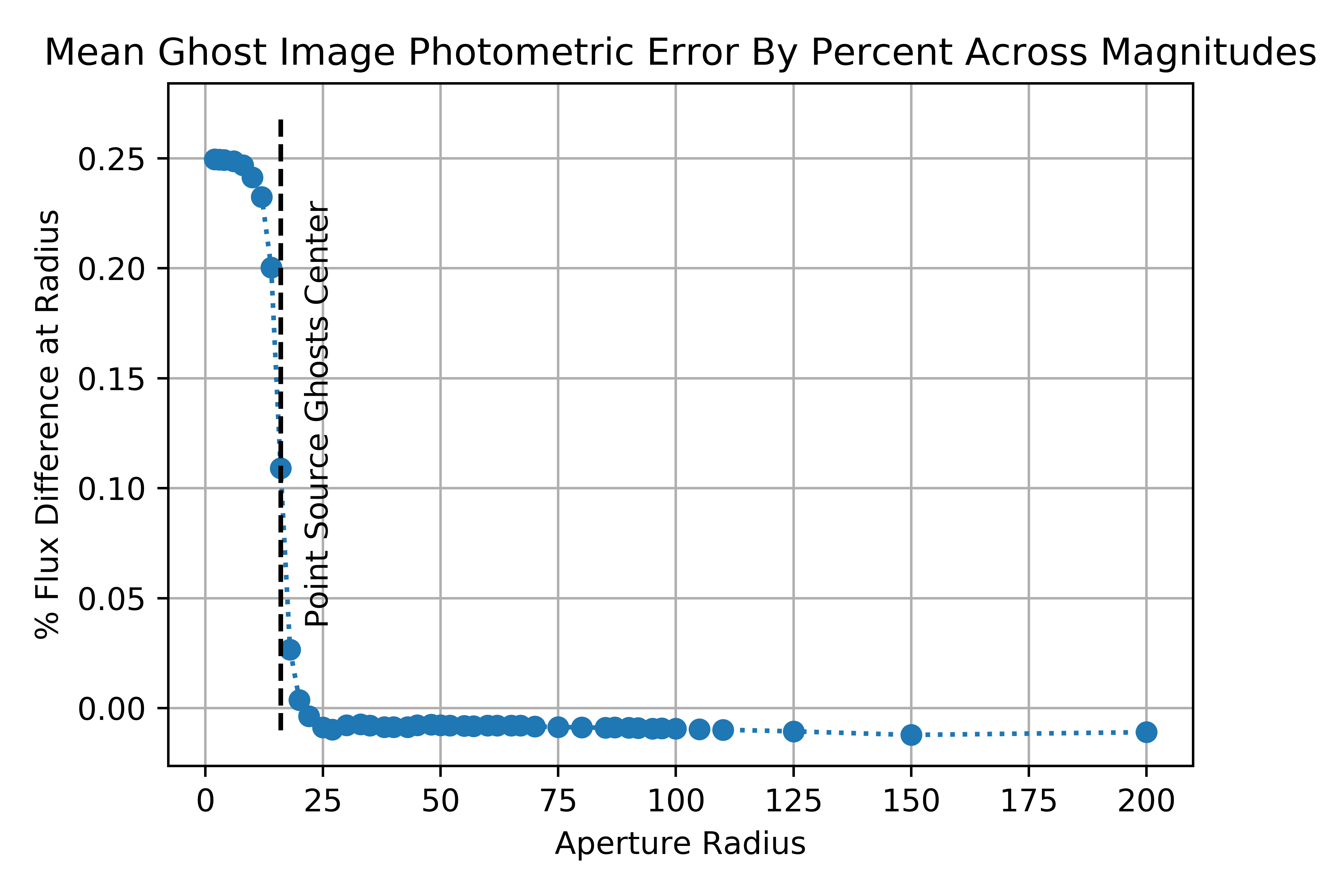}
    \caption{\textbf{Left:} The red dot to the upper right represents the location of the point-source wedge filter ghost image relative to the blue originating point source, at a distance of 63 mas. \textbf{Right:} The photometric error associated with the Wedge ghost in isolation as a function of radius.  Photometric error approaches 0 as the aperture encompasses the affected area.}
    \label{fig:wedgeghost}
\end{figure}

The wedge filter ghost has the highest photometric impact of the predicted ghost images.  Simulated here in isolation of the other ghosts, it exhibits photometric error an order of magnitude large than either of the other ghost images.  Using \textit{Starfinder} we estimate a total photometric impact of 0.267$\pm$0.005\% from the wedge filter ghost. We explore the photometric effects of varying the filter-wedge angle to alter the wedge ghost location in Appendix A. 

\subsubsection{Integrated Ghost Images}

We have presented the estimated photometric effects of the three predicted ghost image types.  We incorporate each of these ghost images in the simulator to analyze their potential effects on photometry. Figure 4 represents the complete ghost model used in the IRIS photometry simulations. 

\begin{figure}[h!]
    \centering
    
    \includegraphics[scale=0.7]{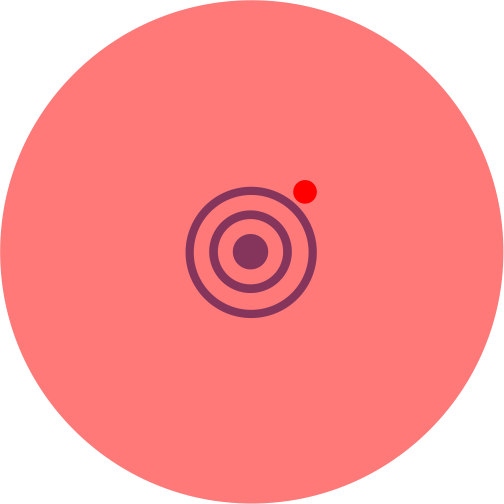}
    \includegraphics[scale=0.6]{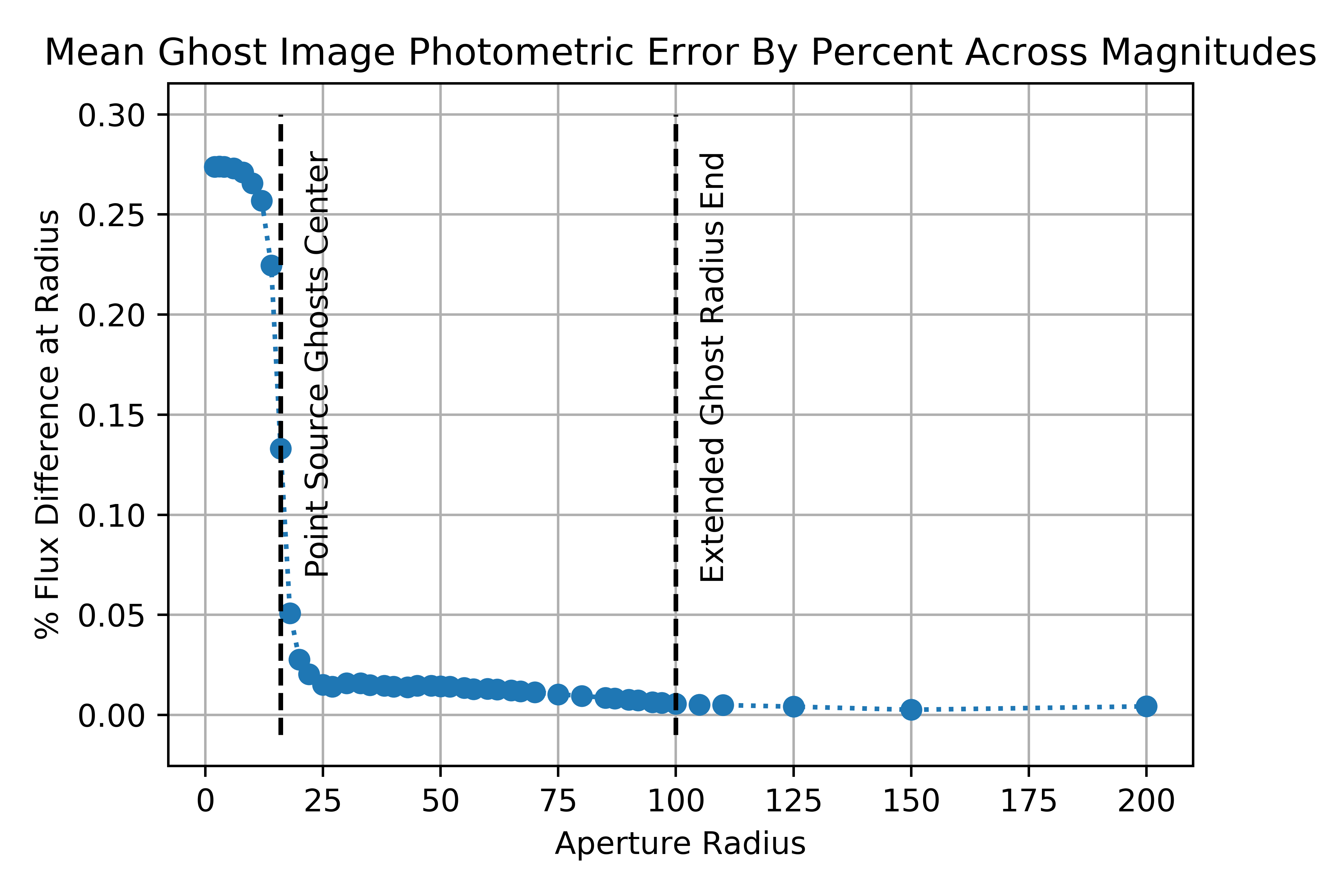}
    \caption{\textbf{Left:} The complete ghosting model including all ghost images, as described in section 2.2. \textbf{Right:} The photometric error associated with the average flux of simulated point sources as a function of radius.  Photometric error approaches 0\% as the aperture encompasses the area affected by each ghost, primarily that of the wedge filters.}
    \label{fig:allghosts}
\end{figure}

As shown in Figures \ref{fig:entranceghost}, \ref{fig:wedgeghost}, and \ref{fig:allghosts}, the estimated photometric error for aperture photometry is a function of radius, and is highly dependent on the location of the ghost images.  Once the aperture used for flux comparison encompasses the ghost image, the photometric error falls to near 0\%, with fluctuations on the order of the Poisson noise of the detector added to the error introduced from the ADC ghost (as shown in Figure \ref{fig:adcghost}). In order to gain a more meaningful understanding of the error introduced we use \textit{Starfinder} to compute flux values per point source simulation, and compare the IRIS data simulation values with those computed from simulated sources not including Poisson noise or ghost images. We bin the magnitude simulations by exposure time appropriate to achieve a high SNR ($>$100) for \textit{Starfinder}.  Figure 5 shows the \textit{Starfinder} results for the predicted ghost image case, where photometric error is consistent per simulation for the high-SNR case at 0.292$\pm$0.005\%.

\begin{figure}[h!]
    \centering
    
    \includegraphics[scale=0.7]{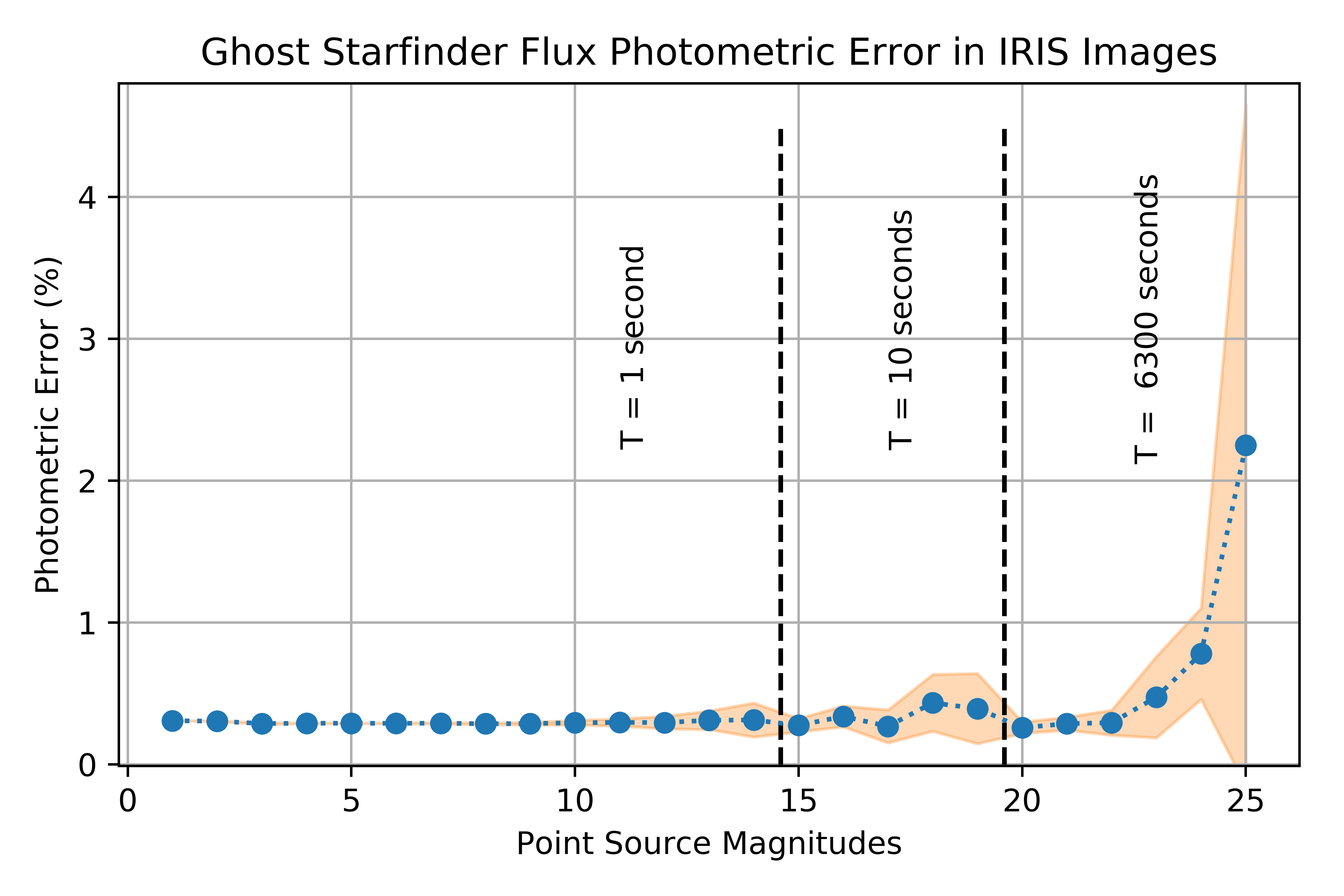}
    \caption{The photometric error associated with the average flux of simulated point sources according to \textit{Starfinder}. Exposure time T marked with corresponding magnitude ranges. Consistency of the estimated photometric error between simulations decreases with SNR, as shown with the increase in associated error (shown in orange).}
    \label{fig:ExpGhosts}
\end{figure}

These ghost image types are included in every simulation conducted with the IRIS simulator.  We characterize the ghost images in terms of photometric error through simulations of IRIS images, in comparison with the calculated flux values corresponding to a given magnitude per filter bandpass. We estimate the photometric error using both aperture photometry and \textit{Starfinder}. Because flux is conserved for each point source simulation, the inclusion of ghost images results in fainter sources until ghost image flux is reintroduced in the photometry for a given magnitude point source. Table \ref{table:1} summarizes the photometric effects of the individual ghosts in isolation as well as the integrated ghost model.

Above, we have shown the photometric error calculations as it pertains to a single point source in the expected ghost image case, and now assess the associated error with varied ghost image characteristics and multiple point sources.
\begin{table}[h!]
\centering

\begin{tabular}{c|c|c}
    \multicolumn{3}{ c }{Ghost Simulation Results}\\
    \hline
     & {Figure Depiction} & {Photometric Error}  \\
     \hline
     \tabcell{Wedge Filter Ghost \\ ($\Delta$m=+6.5)} & \tabcell{\includegraphics[scale=0.5]{WedgeFilterGhostComparisons.png}} & 0.267$\pm$0.005\%   \\
     \hline
     \tabcell{ADC Prism Ghost \\ ($\Delta$m=+5.5/arcsec$^{2}$} & \tabcell{\includegraphics[scale=0.1]{JustADCGhost.png}} & 0.0173$\pm$0.005\% \\
     \hline
     \tabcell{Entrance Window Ghost \\ ($\Delta$m=+9.4/arcsec$^{2}$)} & \tabcell{\includegraphics[scale=0.15]{EntranceWindowClipart.png}} & 0.010$\pm$0.001\% \\
     \hline
     All Ghosts & \tabcell{\includegraphics[scale=0.15]{ExtendedEntranceGhostComparisons.png}} & 0.292$\pm$0.005\% \\
    
\end{tabular}
\caption{Summary of the estimated photometric impact of the IRIS optical system with isolated ghost image simulations for the standard, predicted ghost cases.}
\label{table:1}
\end{table}

\subsection{Ghosts on Single and Multiple Sources}

We assess various orientations of multiple stars in the IRIS field to explore the potentially adverse effects of ghosts on science cases for the predicted ghost image orientations.

\subsubsection{ADC Ghost on a Single Point Source}

\begin{figure}[H]
    \centering
    
    \includegraphics[scale=1.5]{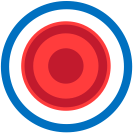}
    \includegraphics[scale=0.4]{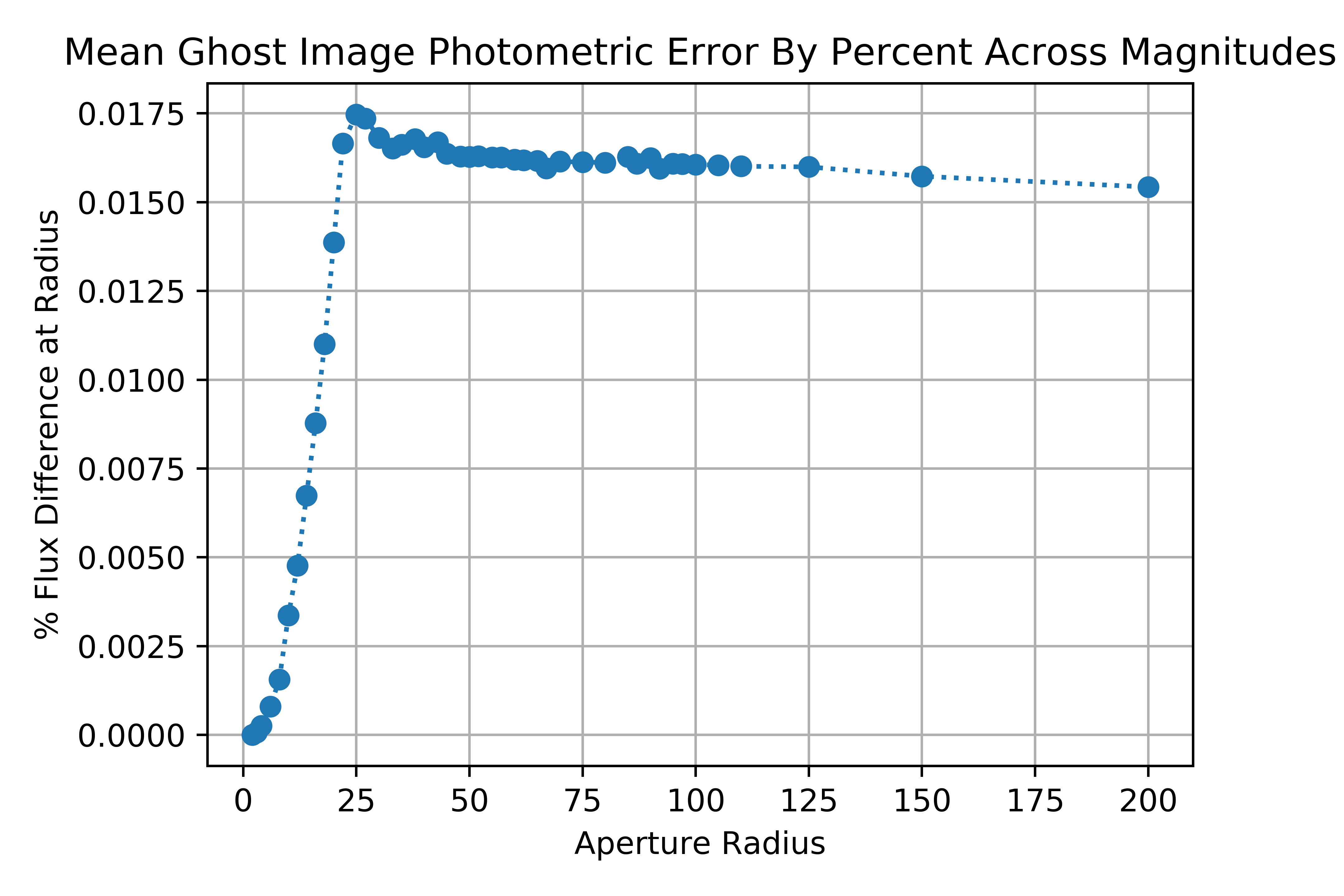}
    \includegraphics[scale=0.4]{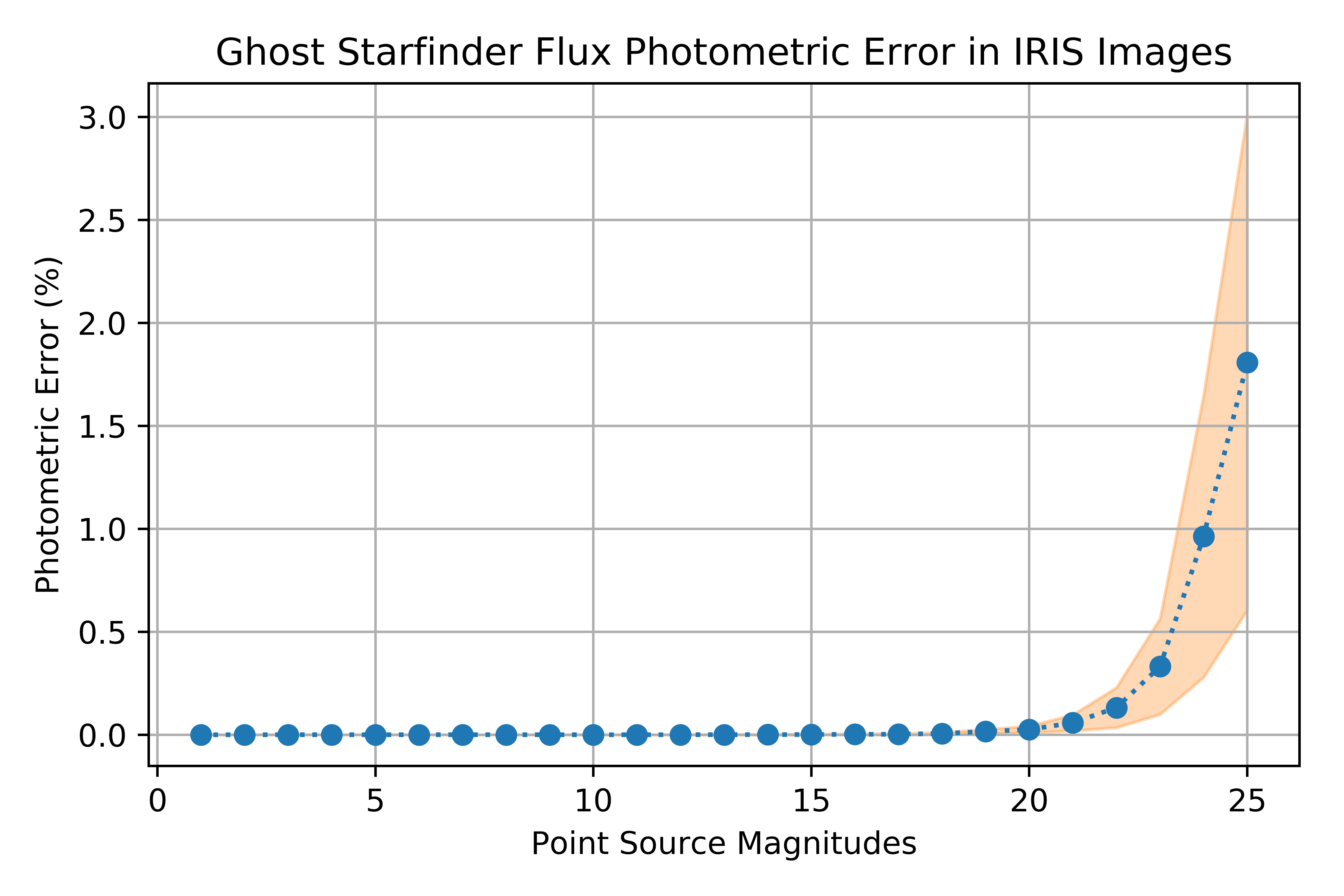}
    \caption{\textbf{Left:} Representation of the simulated point source, with the ADC ghost from a source of equal magnitude projected directly over it. \textbf{Middle:} The photometric error associated with the average flux of simulated point sources as a function of radius. \textbf{Right:} The photometric error associated with each magnitude point source simulation for high SNR as calculated by \textit{Starfinder}.}
    \label{fig:adconlysim}
\end{figure}

We simulate the photometric impact of ADC Prism ghost image were it projected over a science object of equal magnitude ($\Delta$m = 0) to the ghost source. Figure \ref{fig:adconlysim} illustrates the results of this simulation. 

\begin{figure}[H]
    \centering
   
    \includegraphics[scale=1.5]{ADConSourceGhost.png}
    \includegraphics[scale=0.4]{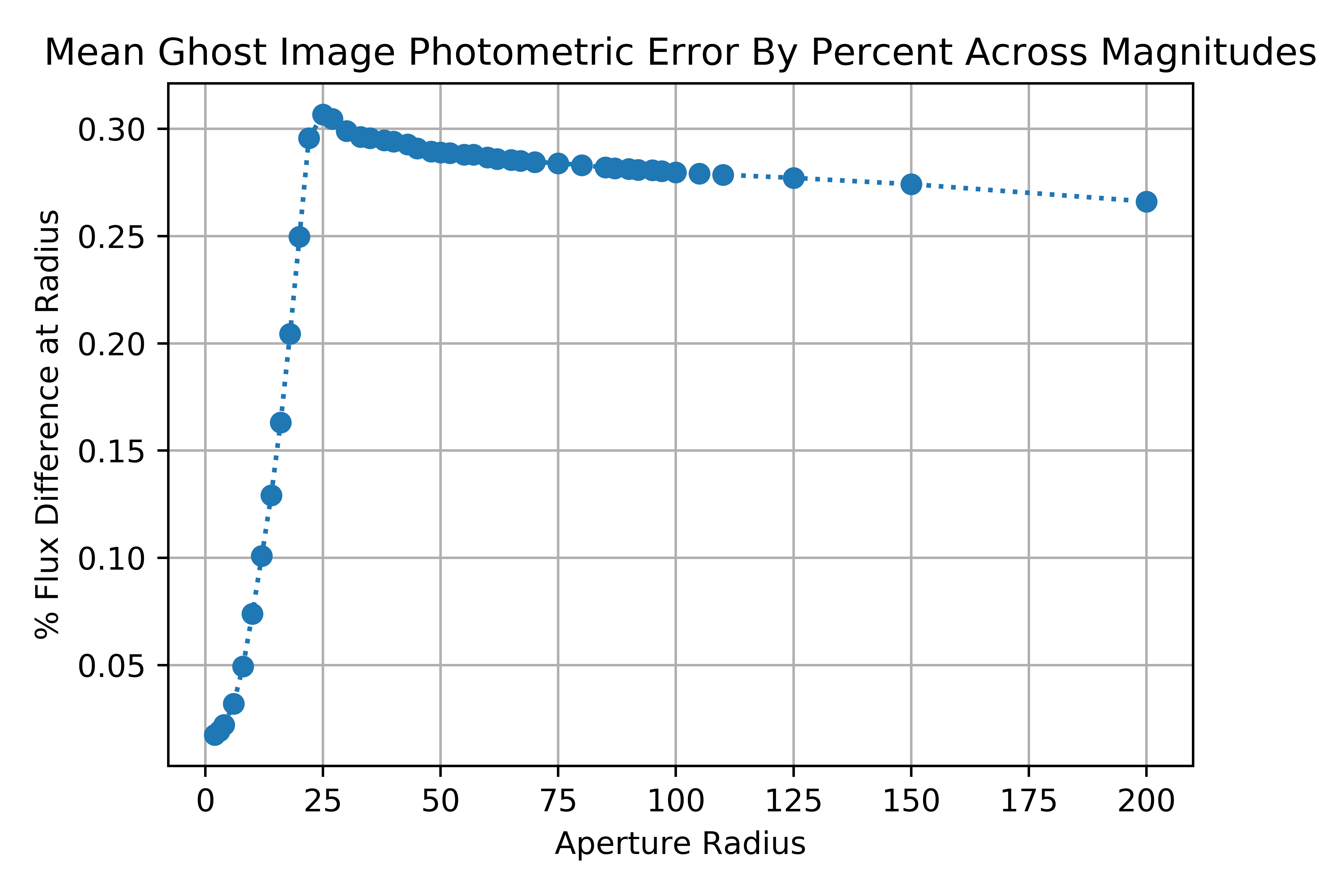}
    \includegraphics[scale=0.4]{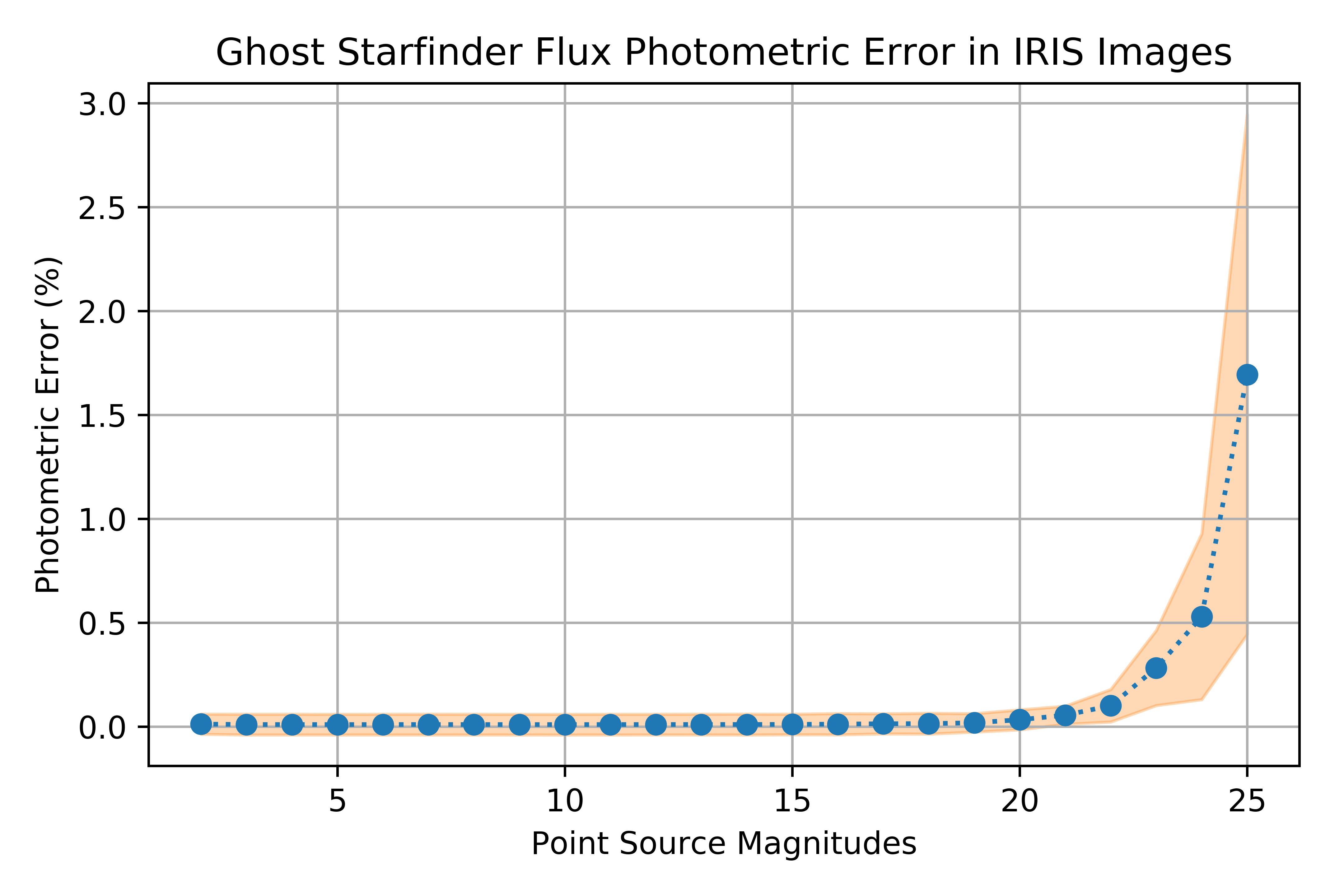}
     \caption{\textbf{Left:} Representation of the simulated point source, with the ADC ghost from a source three magnitudes brighter projected directly over it. \textbf{Right:} The photometric error associated with the average flux of simulated point sources as a function of radius. \textbf{Right:} The photometric error associated with each magnitude point source simulation for high SNR as calculated by \textit{Starfinder}.}
    \label{fig:adcsim}
\end{figure}

For the purpose of solely analyzing the photometric effect of the ADC ghost, this simulated science source does not include its own ghost images.  Because the flux of the ADC ghost does not originate from the science source, photometric error increases as the ghost is encompassed in the aperture radius.  Aperture photometry results in a maximum photometric error of 0.017\% for this case, and \textit{Starfinder} results yield 0.002$\pm$0.003\% photometric error. 
We also assess the photometric impact of the ADC Ghost on a source of $\Delta$m = +3 fainter than the ghost source. The results of this simulation are presented in Figure \ref{fig:adcsim}.

This is a concerning case due to the distance of the ADC prism ghost from its source, meaning this is more likely to occur on potential science sources far from a bright source. Apeture photometry results in a maximum photometric error of 0.306\%.  \textit{Starfinder} results in a photometric error of 0.012$\pm$0.002\%. PSF fitting with \textit{Starfinder} is capable of extracting a much more accurate flux despite the extended ghost image centered over the point source in this case.

\subsubsection{Wedge Ghost on Point Source Cases}

\begin{figure}[H]
    \centering
    
    \includegraphics[scale=0.5]{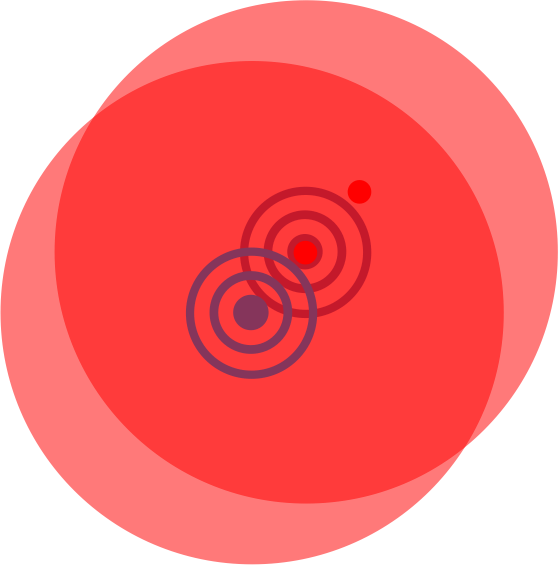}
    \includegraphics[scale=0.4]{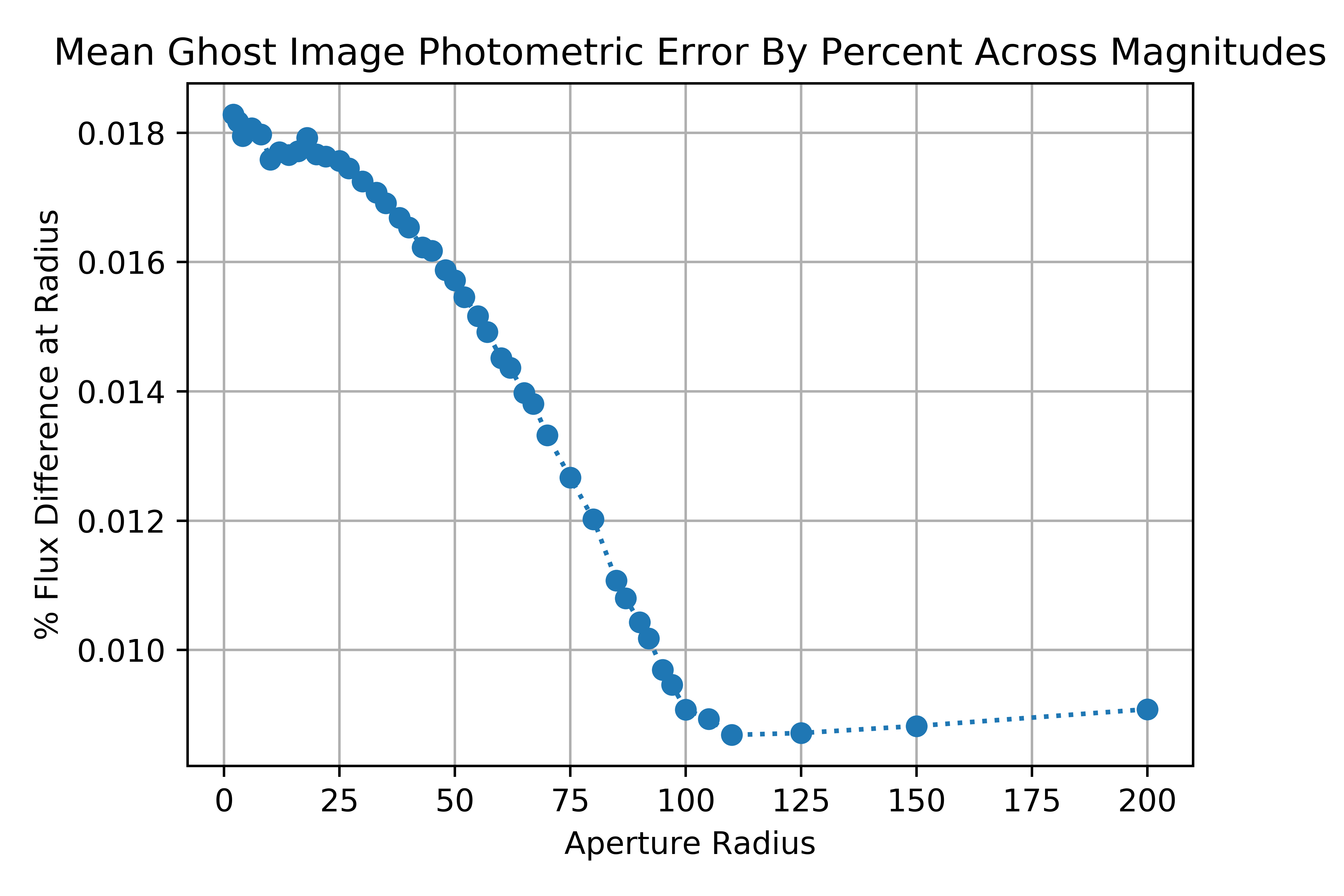}
    \includegraphics[scale=0.4]{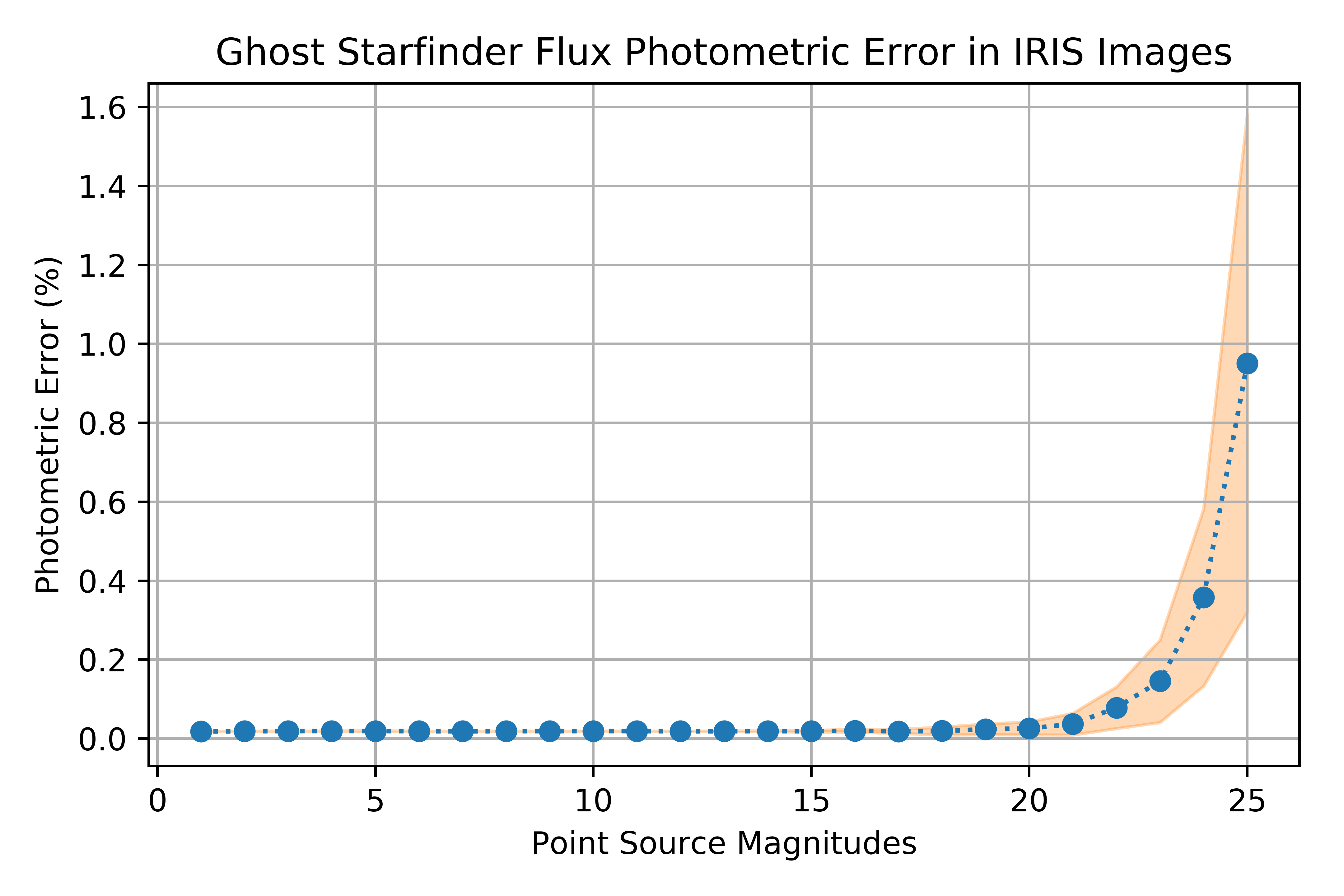}
    \caption{\textbf{Left:} Representation of the simulated point source, with the wedge ghost from a source of equal magnitude projected directly over it. \textbf{Right:} The photometric error associated with the average flux of simulated point sources as a function of radius. \textbf{Right:} The photometric error associated with each magnitude point source simulation for high SNR as calculated by \textit{Starfinder}.}
    \label{fig:wedgesim}
\end{figure}

We simulate the photometric impact of wedge filter ghost image were it projected over a science object of equal magnitude ($\Delta$m = 0) to the ghost source. Figure \ref{fig:wedgesim} illustrates the results of this simulation.

\begin{figure}[H]
    \centering
    
    \includegraphics[scale=0.5]{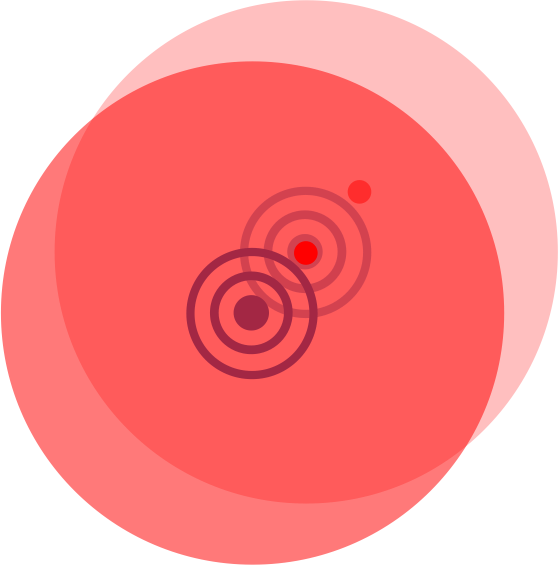}
    \includegraphics[scale=0.4]{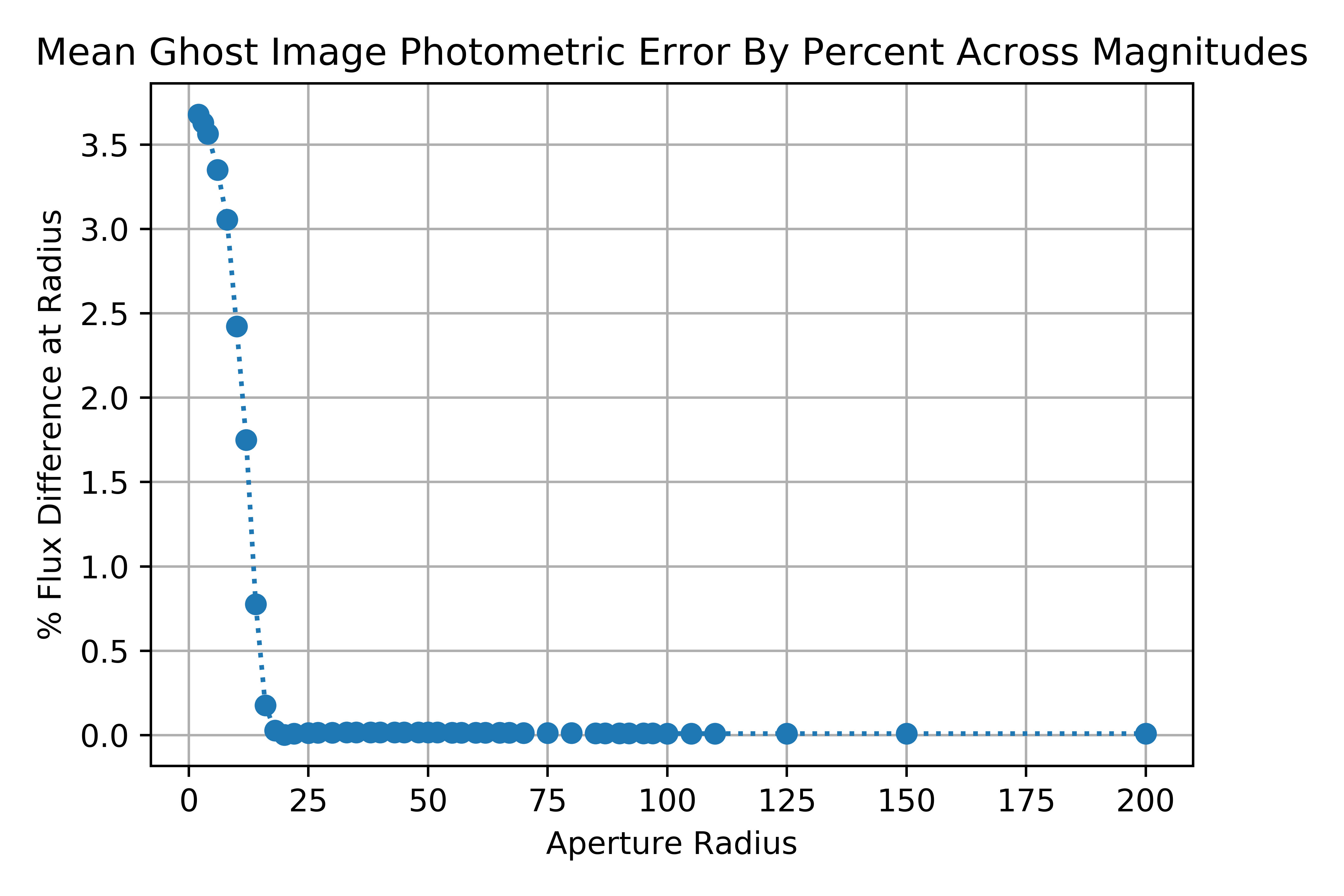}
    \includegraphics[scale=0.4]{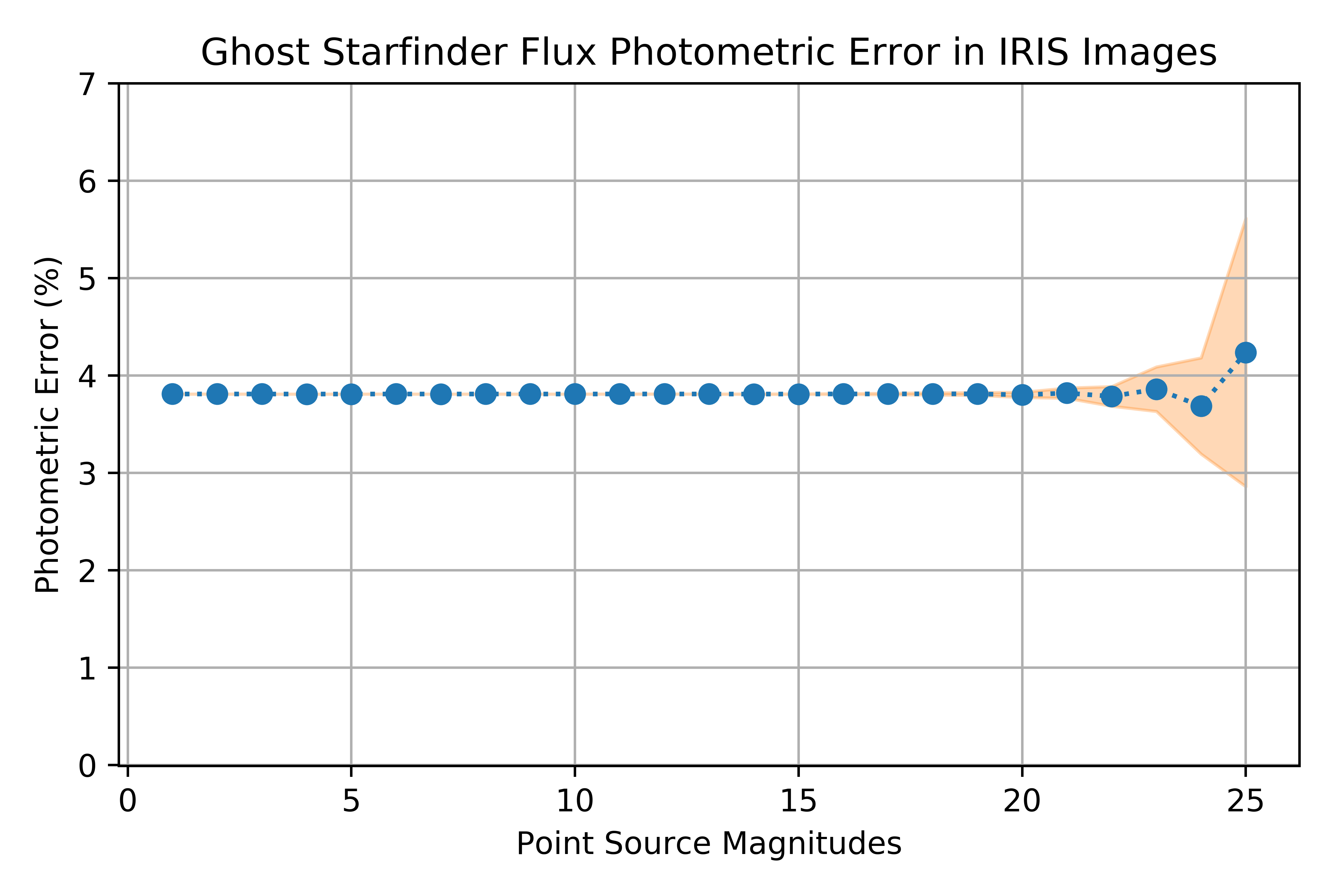}
    \caption{\textbf{Left:} Representation of the simulated point source, with the wedge ghost from a source three magnitudes brighter projected directly over it. \textbf{Right:} The photometric error associated with the average flux of simulated point sources as a function of radius. \textbf{Right:} The photometric error associated with each magnitude point source simulation for high SNR as calculated by \textit{Starfinder}.}
    \label{fig:wedgedimsim}
\end{figure}

Apeture photometry results in a maximum photometric error of 0.018\%.  \textit{Starfinder} results in a photometric error of 0.019$\pm$0.001\%. This case results in slightly decreased photometric error, as the wedge filter from the secondary source is placed directly on top of the science source, reintroducing the flux lost from the its wedge filter ghost. We also simulate the photometric impact of wedge filter ghost image were it projected over a science object of 3 magnitudes fainter ($\Delta$m = +3) than the ghost source. Figure \ref{fig:wedgedimsim} illustrates the results of this simulation.

This is an extreme case in which the science source is coaligned with the wedge ghost of a source 3 magnitudes brighter, resulting in high photometric error.  This is the worst-case scenario out of the simulated science cases, with a maximum aperture photometry error of 3.680\% and \textit{Starfinder} results yielding 3.810$\pm$0.002\% for this case.

\subsubsection{Crowded Field}
\begin{figure}[H]
    \centering
    
    \includegraphics[scale=0.4]{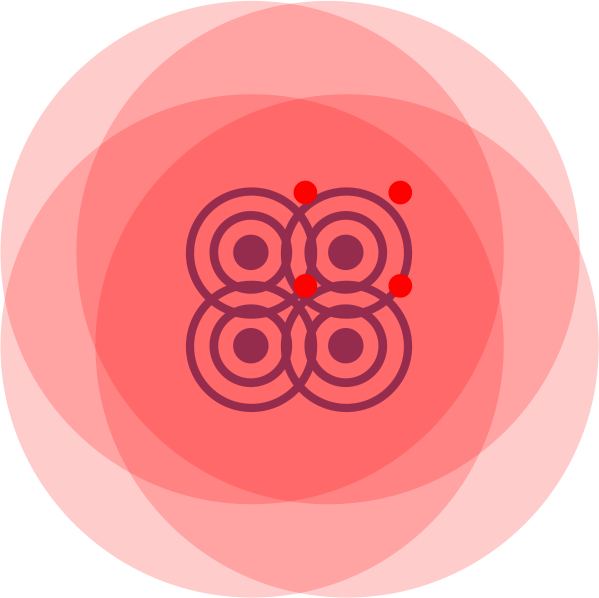}
    \includegraphics[scale=0.4]{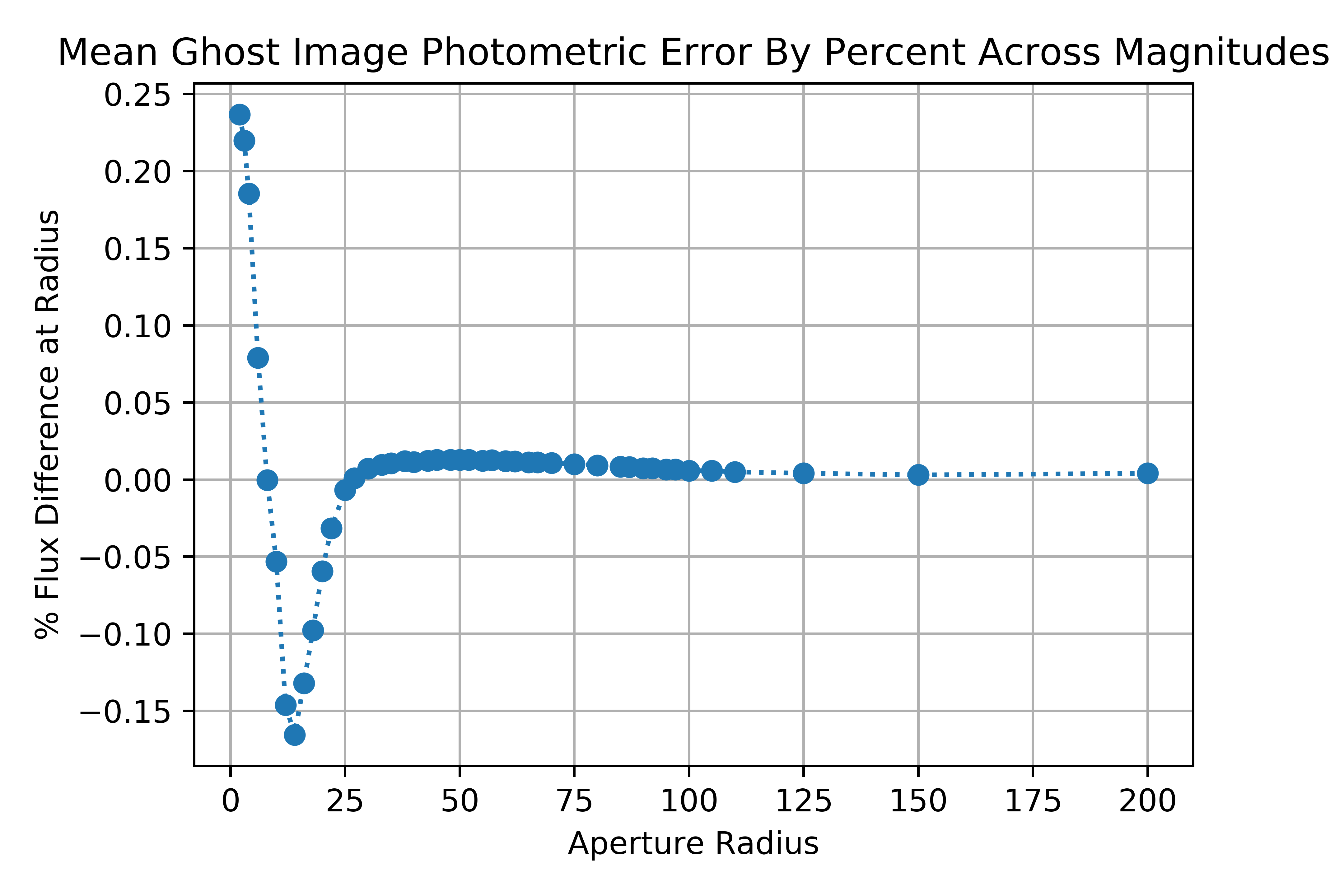}
    \includegraphics[scale=0.4]{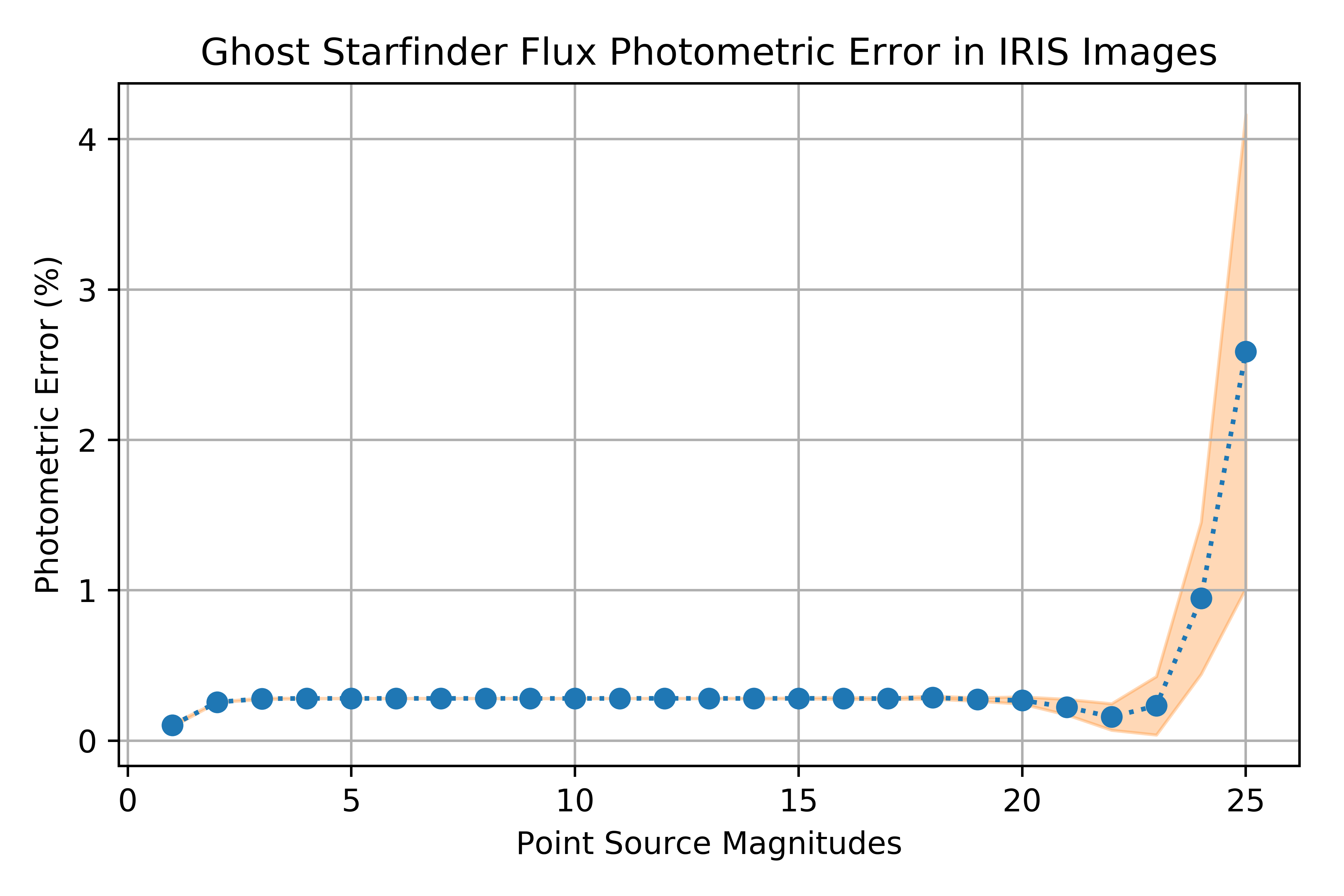}
    \caption{\textbf{Left:} Representation of simulated science case. \textbf{Right:} The photometric error associated with the average flux of simulated point sources as a function of radius. \textbf{Right:} The photometric error associated with each magnitude point source simulation for high SNR as calculated by \textit{Starfinder}.}
    \label{fig:crowdedfield}
\end{figure}

We simulate the photometric impact of a crowded field of similarly bright sources with ghost images, at a distance of 60 mas (15 pixels at 4 mas/pixel scale) from each other, placing wedge filter ghosts close to different sources. Figure \ref{fig:crowdedfield} illustrates the results of this simulation.
Here we simulate the case of extremely close sources of similar magnitude. This is useful to see how the entrance window ghost may behave in a crowded field, as the flux from multiple sources is spread over others. The results of the simulation show that the wedge filter ghost still dominates; maximum photometric error from aperture photometry is shown to be 0.236\%, and \textit{Starfinder} results show an average of 0.280$\pm$0.003\%.

\subsubsection{Summary of Multiple-Source Ghost Impact}

We summarize the calculated photometric impact of the optical system on the IRIS Imager including various ghost image simulations in Table \ref{table:3}.

\begin{table}[H]
\centering

\begin{tabular}{c|c|c|c}
    \multicolumn{4}{ c }{Multiple Star Ghost Simulation Results}\\
    \hline
     & {Figure Depiction} & \tabcell{$\Delta$m of Second \\ Source} & {Photometric Error} \\
     \hline
     \tabcell{ADC Prism Ghost \\ on Science Source} & \tabcell{\includegraphics[scale=0.5]{ADConSourceGhost.png}} & 0 & 0.002$\pm$0.003  \\
     \hline
     \tabcell{ADC Prism Ghost \\ on Dim Science Source} & \tabcell{\includegraphics[scale=0.5]{ADConSourceGhost.png}} & +3 & 0.012$\pm$0.002\% \\
     \hline
     \tabcell{Wedge Filter Ghost \\ on Science Source} & \tabcell{\includegraphics[scale=0.15]{EqSourceWithGhost.png}} & 0 & 0.019$\pm$0.001\%  \\
     \hline
     \tabcell{Wedge Filter Ghost \\ on Dim Science Source} & \tabcell{\includegraphics[scale=0.15]{DimSourceWithGhost.png}} & +3 & 3.810$\pm$0.002\% \\
     \hline
     \tabcell{Four Sources, Distance \\ of D=15p} & \tabcell{\includegraphics[scale=0.15]{FourSourcewGhosts.png}} & 0 & 0.280$\pm$0.003\%  \\
    
\end{tabular}
\caption{Science case simulations assuming multiple sources causing various ghost image scenarios.}
\label{table:3}
\end{table}

\section{Photometric Precision}

The inclusion of ghost images in IRIS simulations results in an increase to the photometric error. Aperture photometry informs how the ghost images effect the photometric error of IRIS images as a function of radius, while using \textit{Starfinder} yields the total flux loss using PSF-modeling. These simulations include all current known contributions to the IRIS imager and we find a relative photometric error of 0.292$\pm$0.005\% from ghosting. 

In terms of relative photometric precision, IRIS is limited by multiple noise sources of varying brightness.  For brighter point source magnitudes IRIS is limited by Poisson noise from the source(s) in the FoV, whereas at fainter magnitudes it is limited by the instrument noise floor. In the IRIS simulator, we assume the following sources of noise in the IRIS imager: readnoise of 5\ $e^-$, dark current of 0.002\ $e^-$, and Poisson noise from sources, background (telescope, instrument, and atmospheric), and dark current\cite{Wright1,Wright2,Wright3}.

We note that systematic error sources are not included in the simulation. Additionally, we use a single PSF per bandpass for these simulations, which does not represent a fair comparison to real observations. In future simulations we will investigate varying the PSF per observational sequence to continue to explore the IRIS imager photometric precision. The IRIS simulator determines the signal-to-noise of a given magnitude point source, using
\begin{equation}
    SNR = \frac{F_* * \sqrt{T}}{\sqrt{F_* + \sigma_{RN} + \sigma_{DN} +\sigma_{B}}}
    \label{eq:snreq}
\end{equation}
where $F_*$ is the IRIS image flux, T is integration time, $\sigma_{RN}$is the read noise, $\sigma_{DN}$ is the dark current noise, and $\sigma_{B}$ is the background noise. Using Equation \ref{eq:snreq} we select appropriate exposure times to yield high SNR values ($>$100) for the magnitudes 15, 20, and 25 in Table \ref{table:mags}. These parameters are used to simulate 25 single frames, and then we use \textit{Starfinder} in Y (0.97 - 1.07 $\mu$m) and K (2 - 2.4 $\mu$m) bandpasses for PSF-modeling to determine the photometric precision per observational sequence.

\begin{table}[H]
    \centering
    
    \begin{tabular}{c|c|c|c|c|c|c}
         \multirow{2}{*}{Bandpass} & \multirow{2}{*}{\tabcell{Integration\\Time (Seconds)}} &  \multirow{2}{*}{Magnitude} & \multirow{2}{*}{\tabcell{Maximum SNR\\(Single Frame)}} &
         \multicolumn{3}{c}{\tabcell{Photometric Precision (\%) \\ \hline}} \\
         && & & \textit{Starfinder} & \tabcell{Aperture $R_1$} & \tabcell{Aperture $R_2$} \\
         \hline
         \multirow{4}{*}{K} & 1 & 15 & 137.8  & 0.1433 & 0.1483 & 0.1435 \\
         \cline{2-7}
         &60 & 20 & 249.7 & 0.1938 & 0.0988 & 0.0703 \\
         \cline{2-7}
         &300 & 20 & 238.8 & 0.1189 & 0.0741 & 0.0988 \\
         \cline{2-7}
         & 900 & 25 & 40.96 & 3.079 & 3.402 & 8.518 \\
         \hline
         \multirow{4}{*}{Y} & 1 & 15 & 322.3 & 0.1451 & 0.0889 & 0.0728 \\
         \cline{2-7}
         &60 & 20 & 106.7 & 0.2793 & 0.1892 & 0.3431 \\
         \cline{2-7}
         &300 & 20 & 558.3 & 0.08572 & 0.0654 & 0.0571 \\
         \cline{2-7}
         &900 & 25 & 96.57 & 0.4930 & 0.3884 & 0.8275 \\
    \end{tabular}
    \label{table:mags}
    \caption{Estimated photometric precision values from \textit{Starfinder} and aperture photometry varying source magnitudes, exposure time, and bandpasses. For aperture photometry we use the two apertures $r_1$ and $r_2$, which we define in terms of the full width half maximum (FWHM) of the PSF: $r_1$ is the PSF FWHM, and $r_2$ is the PSF FWHM multiplied by a factor of 3.}
\end{table}

\begin{figure}[h!]
    \centering
    
    \includegraphics[scale=0.5]{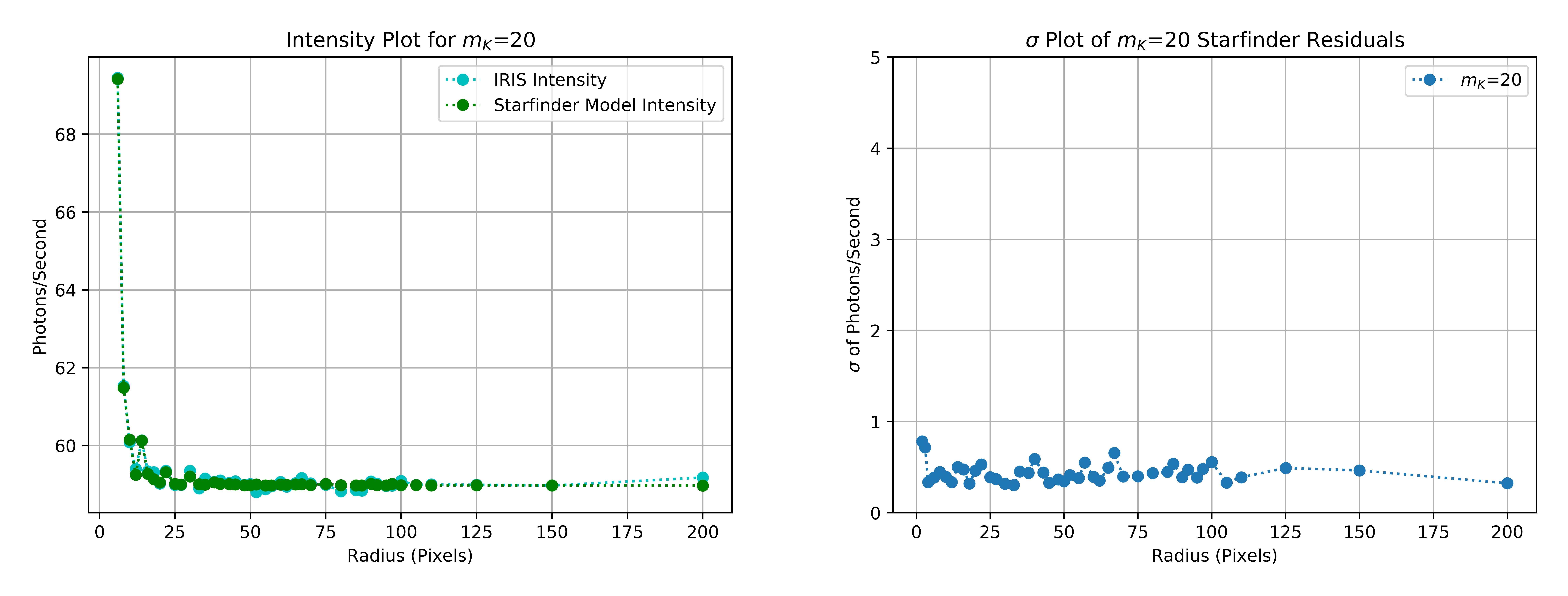}
    \caption{\textbf{Left:} $\sigma$ of the residual intensity across the 25 simulated frames for multiple sources, utilizing \textit{Starfinder}'s ability to model each source to produce a residual image. \textbf{Right:} Intensity plot of the simulated $m_K$=20 source (blue), overplotted with the \textit{Starfinder} model (green).}
    \label{fig:sigmares}
\end{figure}

We estimate the photometric precision of the IRIS imager by analyzing the values and residual images returned by \textit{Starfinder} for each observational sequence (i.e., integration time, bandpass) simulation. \textit{Starfinder} produces a model source and residual image that  subtracts the model from the input image. We then analyze the intensity and standard deviation ($\sigma$) as a function of radius for both the simulated point source image and the residual image, as illustrated in Figure \ref{fig:sigmares}. For an ideal observational sequence, a 1.5 hour observation of K=20 mag point source can theoretically achieve a photometric precision $<$1\%, and 6 hour observation of K=25 mag point source can reach a photometric precision of $<$3\%, see Table \ref{table:mags}. At shorter wavelengths, we achieve better photometric precision given the lower background levels and reach a photometric precision of $<$1\% for Y=25 mag point source.

In the future we plan to introduce more photometric effects into the simulator which will influence IRIS photometric precision.  We will analyze the photometric effects of varying the PSF across the imager field of view, as well as include more updated information for the atmospheric and instrument errors estimated in the IRIS optical design. Most importantly, we will vary the PSF per simulation in order to approximate atmospheric effects, which will increase IRIS overall photometric error. We will also investigate the effects of ghost images for the IRIS Integral Field Spectrograph. IRIS has a pick-off mirror for the IFS at the center of the imager field of view\cite{Larkin4}, making it vulnerable to potential ghost images from sources seen by the imager. We will also expand the wavelength range of these simulations and analyze how the photometric effects change in varying bandpasses.

\section{Summary}

We have presented the results of ghost image analysis by simulating point sources with the IRIS imager and calculating the associated photometric error using PSF-fitting and aperture photometry. We explored the photometric impact of individual ghost image types (wedge filter, ADC, entrance window) from the IRIS optical design, and find that ghosts from the wedge-shaped filters have the highest associated photometric error. We investigated various multiple-point-source scenarios where ghost images could adversely affect photometry, and find that ghosts generally contriburte negligible effects (less than 1\% photometric error) except in extreme cases where ghosts coalign with a star of $\Delta$m$>$2 fainter than the ghost source. We report the impact on IRIS's photometric accuracy from ghosting to to be 0.292$\pm$0.005\% for a single point source.  We report preliminary values for IRIS's photometric precision for a single point source at varying magnitudes and integration time, with a potential for $<$1$\%$ precision at $m_Y$=25 and $\sim$ 3\% at $m_K$=25, and anticipate further development of IRIS's predicted photometric specifications in parallel with IRIS's advancing design. We plan further development of these simulations with the addition of systematic errors and PSF variability, which will better predict the photometric performance of IRIS.

%%%%%%%%%%%%%%%%%%%%%%%%%%%%%%%%%%%%%%%%%%%%%%%%%%%%%%%%%%%%%
\acknowledgments     %>>>> equivalent to \section*{ACKNOWLEDGMENTS}       
 
The TMT Project gratefully acknowledges the support of the TMT collaborating institutions. They are the California Institute of Technology, the University of California, the National Astronomical Observatory of Japan, the National Astronomical Observatories of China and their consortium partners, the Department of Science and Technology of India and their supported institutes, and the National Research Council of Canada. This work was supported as well by the Gordon and Betty Moore Foundation, the Canada Foundation for Innovation, the Ontario Ministry of Research and Innovation, the Natural Sciences and Engineering Research Council of Canada, the British Columbia Knowledge Development Fund, the Association of Canadian Universities for Research in Astronomy (ACURA), the Association of Universities for Research in Astronomy (AURA), the U.S. National Science Foundation, the National Institutes of Natural Sciences of Japan, and the Department of Atomic Energy of India.

%%%%%%%%%%%%%%%%%%%%%%%%%%%%%%%%%%%%%%%%%%%%%%%%%%%%%%%%%%%%%
%%%%% References %%%%%
% Cherenkov instruments (veritas,hawk, FAMOUS, cosmic house, nicholas law)
% SETI (general, optical/nir)
% structural (popko, stiffness)
% fresnel

\bibliographystyle{spiebib}
\bibliography{article.bbl}   %>>>> bibliography data in report.bib

\appendix

\section{Wedge-Filter Angle Ghost Variations}

We assess the photometric impact of ghost images as we vary the wedge-angle associated with the IRIS wedge-filters, thus changing the distance of the wedge filter ghost.

\begin{figure}[h!]
    \centering
    
    \includegraphics[scale=0.5]{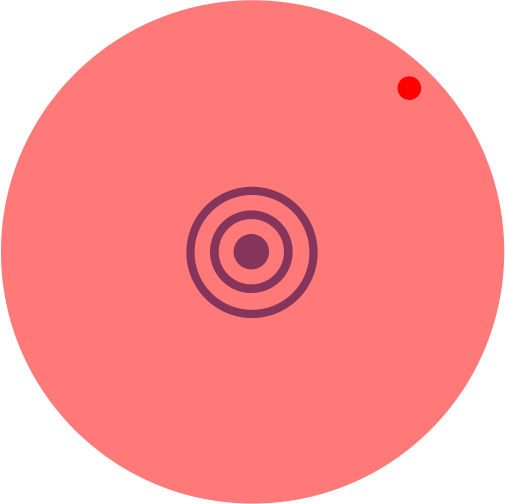}
    \includegraphics[scale=0.4]{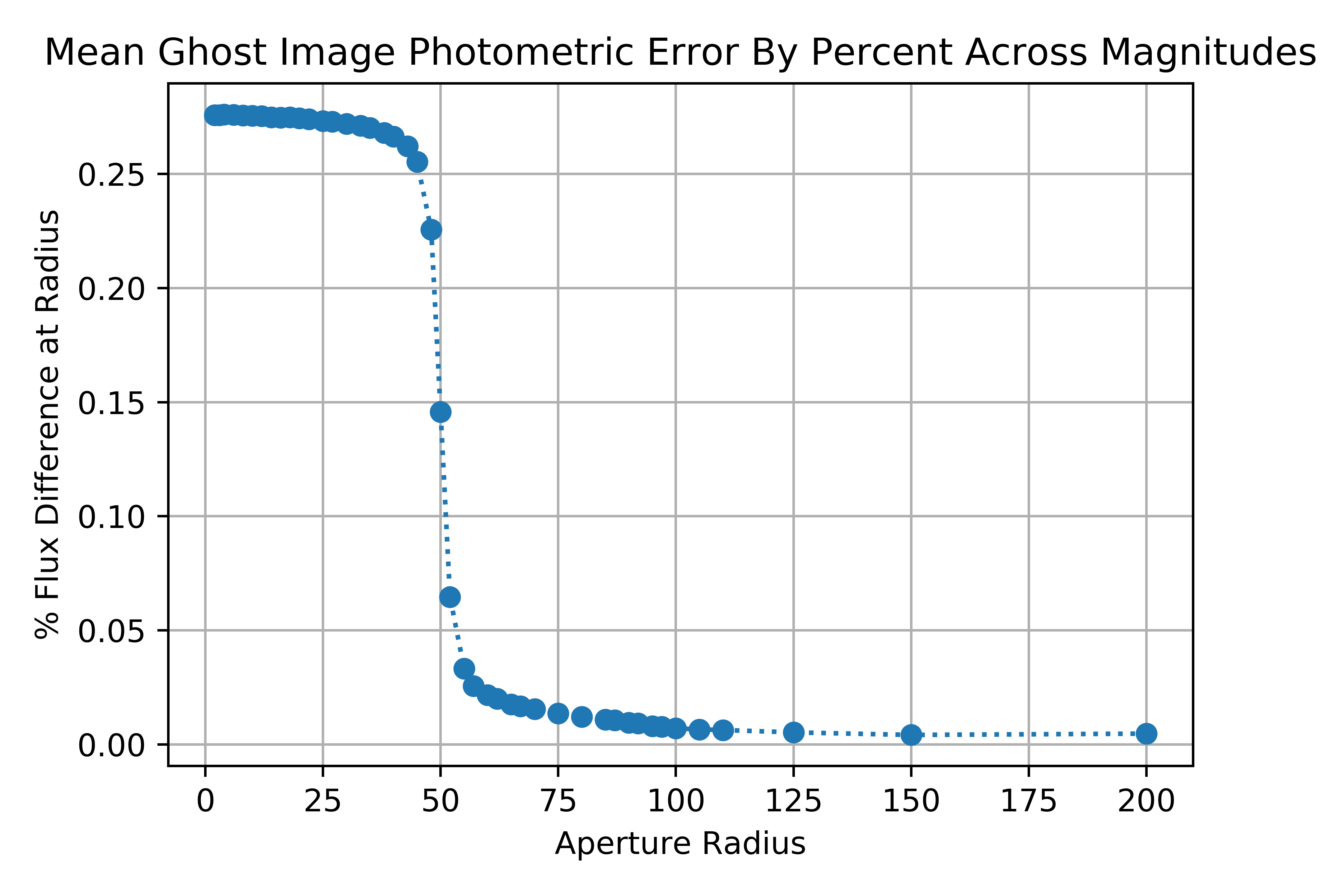}
    \includegraphics[scale=0.4]{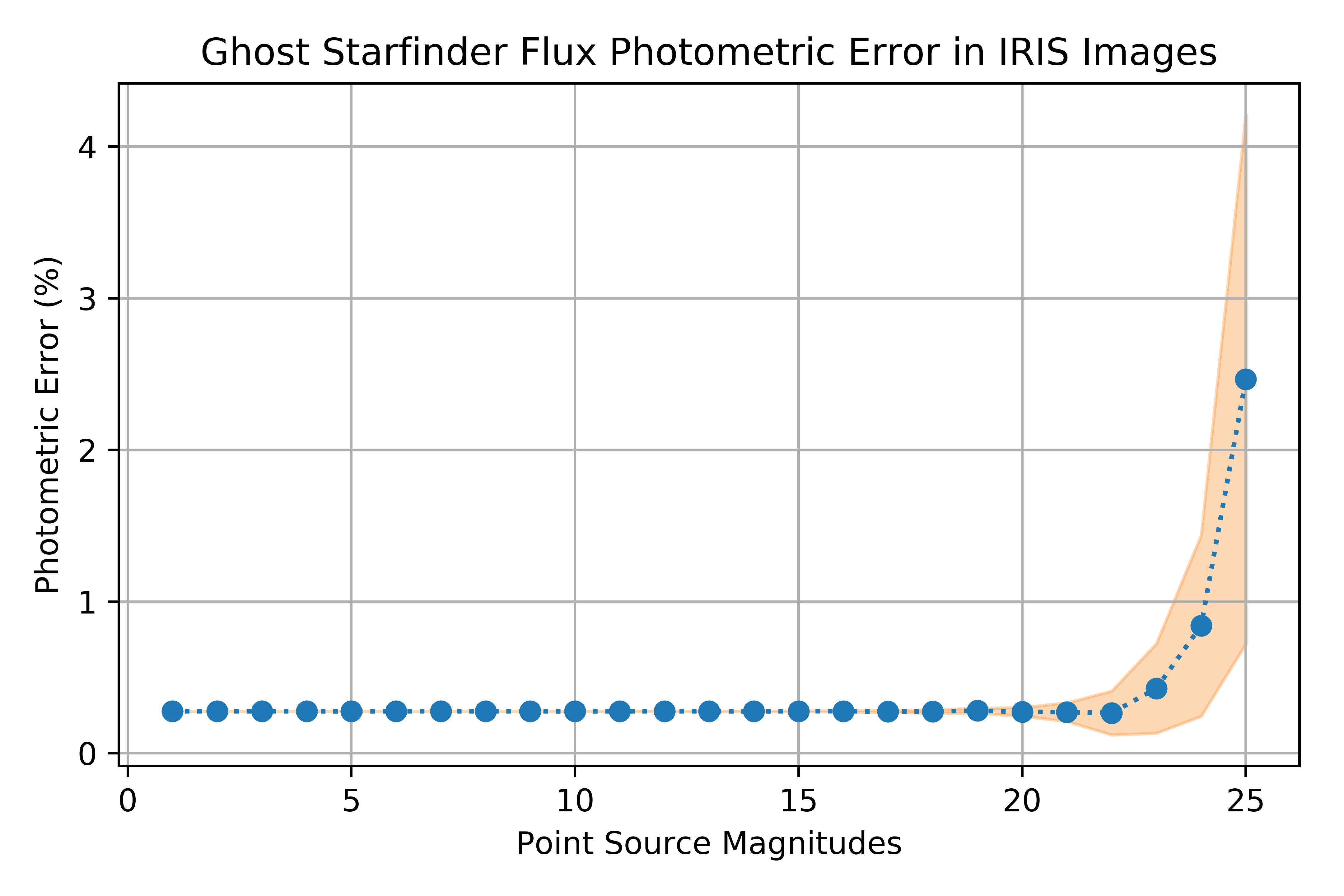}
    \caption{\textbf{Left:} Representation of simulated wedge ghost distance, d=50 pixels. \textbf{Middle:} The photometric error associated with the average flux of simulated point sources as a function of radius.  Photometric error approaches 0 after encompassing the wedge filter ghost. \textbf{Right:} The photometric error associated with each magnitude point source simulation for high SNR as calculated by \textit{Starfinder}.}
    \label{fig:r50ghost}
\end{figure}

We simulate the photometric impact of the wedge filter ghost image assuming a larger filter-wedge angle, resulting in a distance of 0.2 arcseconds (50 pixels). Figure \ref{fig:r50ghost} illustrates the results of this simulation. 
Moving the wedge filter ghost to a location more distant from its source increases the total photometric error over distance, although the maximum photometric error according to aperture photometry remains consistent at 0.276\%.  \textit{Starfinder} results also show high-SNR average photometric error of 0.276$\pm$0.001\%.  This case is less preferable than the standard ghost image case not only due to the increased total photometric error, but also because increased distance from the ghost source increases the likelihood of this ghost being projected over a potential science source.  The photometric impact of such a scenario is explored in section 2.2.2.

\begin{figure}[h!]
    \centering
    
    \includegraphics[scale=0.5]{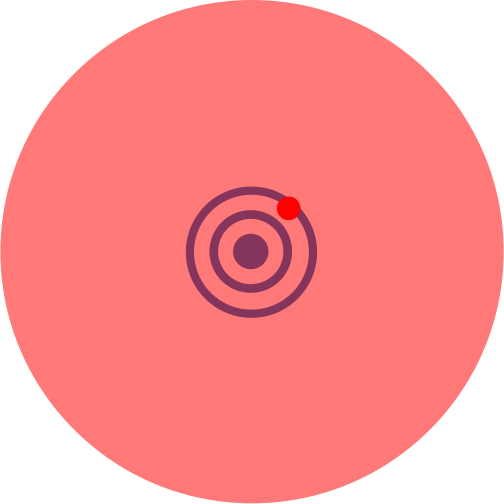}
    \includegraphics[scale=0.4]{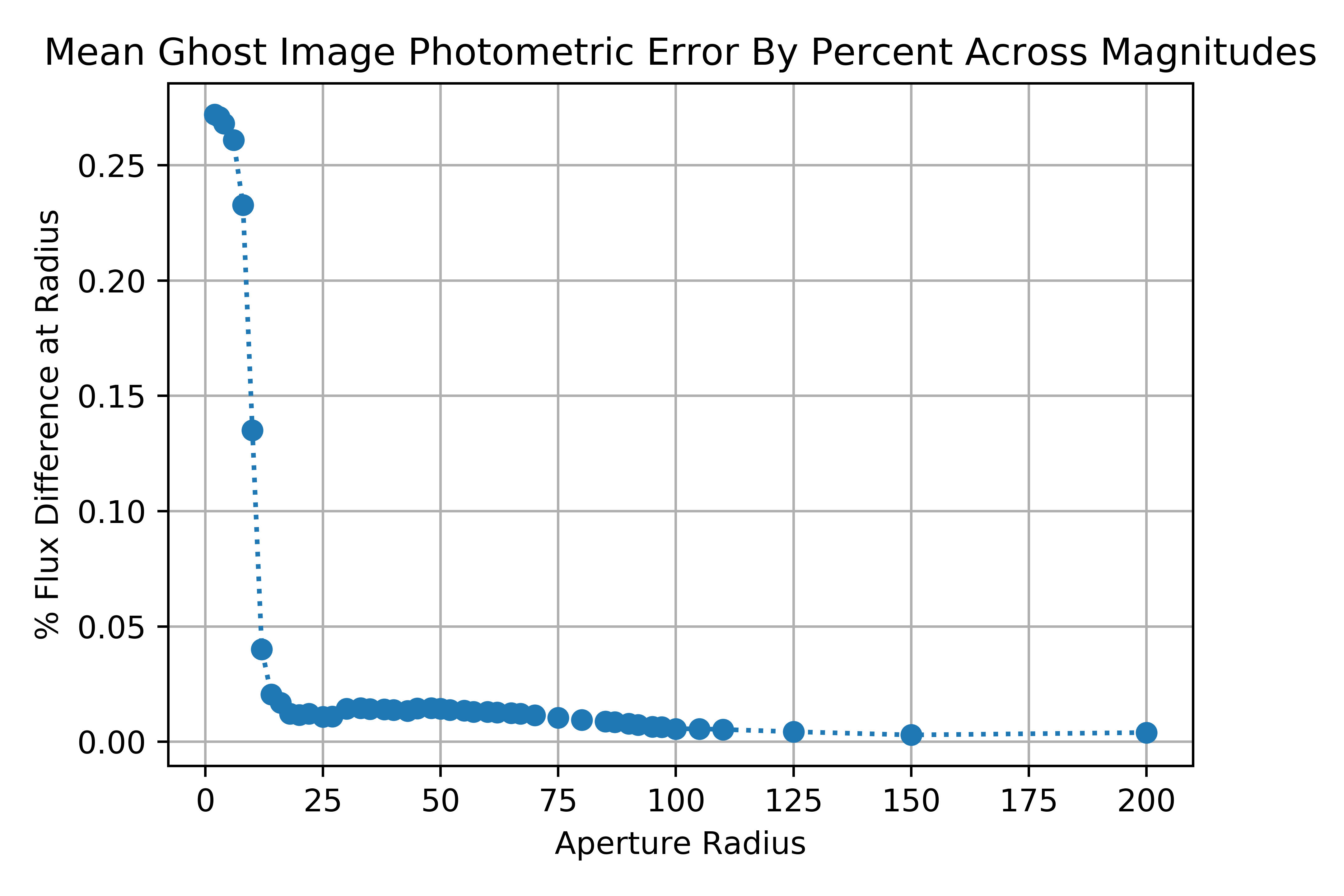}
    \includegraphics[scale=0.4]{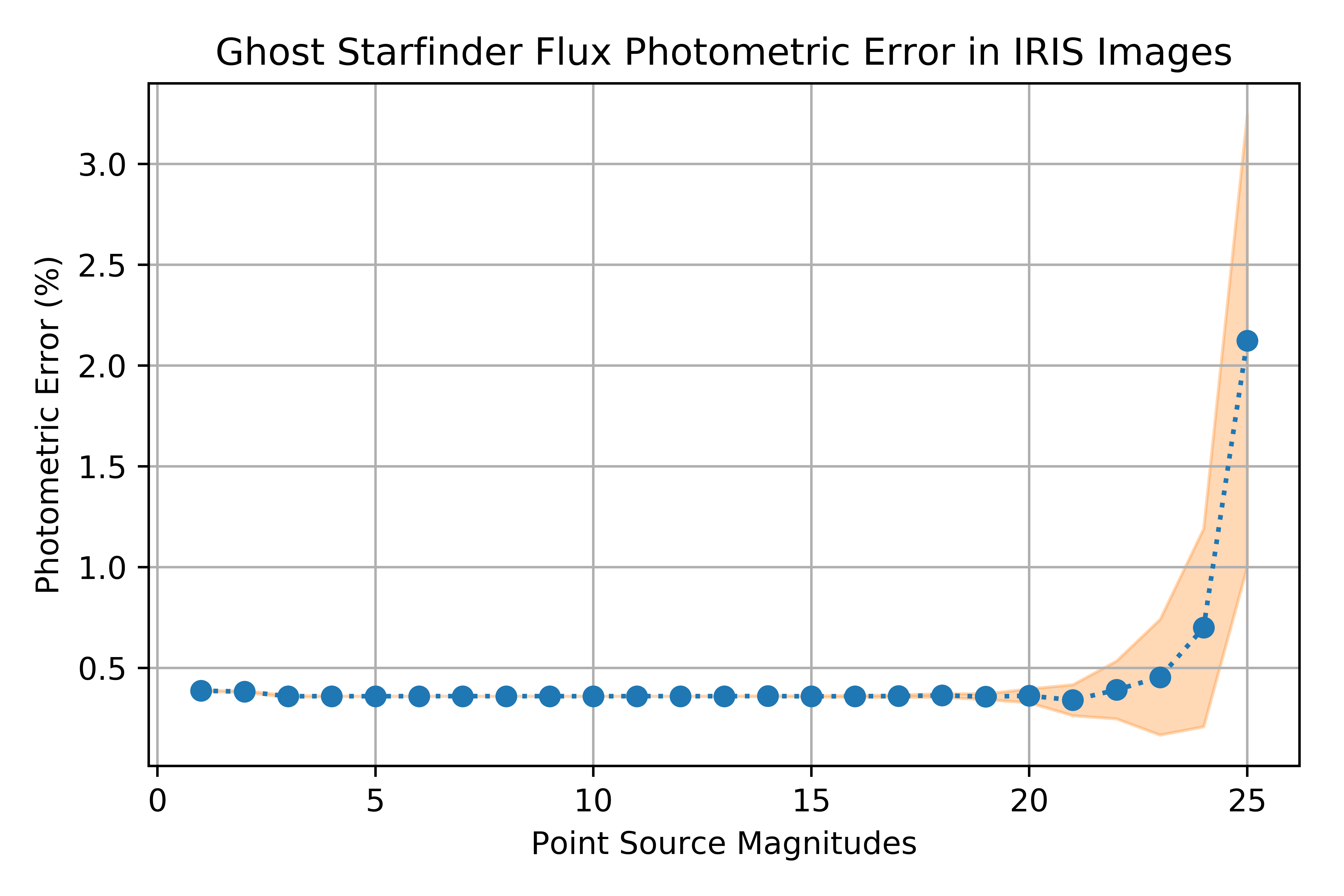}
    \caption{\textbf{Left:} Representation of simulated wedge ghost distance, d=10 pixels. \textbf{Middle:} The photometric error associated with the average flux of simulated point sources as a function of radius.  Photometric error approaches 0 after encompassing the wedge filter ghost. \textbf{Right:} The photometric error associated with each magnitude point source simulation for high SNR as calculated by \textit{Starfinder}.}
    \label{fig:r10ghost}
\end{figure}

We simulate the photometric impact of the wedge filter ghost image assuming a smaller filter-wedge angle, resulting in a distance of 40 mas (10 pixels). Figure \ref{fig:r10ghost} illustrates the results of this simulation. 

The wedge filter ghost closer to the ghost source results in less overall photometric error according to aperture photometry, which yields a maximum of 0.272\%. \textit{Starfinder} results yield a 0.364$\pm$0.008\% photometric error associated with this case.

\begin{figure}[h!]
    \centering
    
    \includegraphics[scale=0.5]{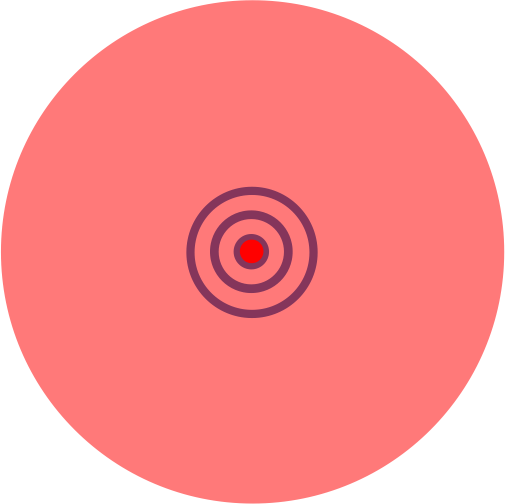}
    \includegraphics[scale=0.4]{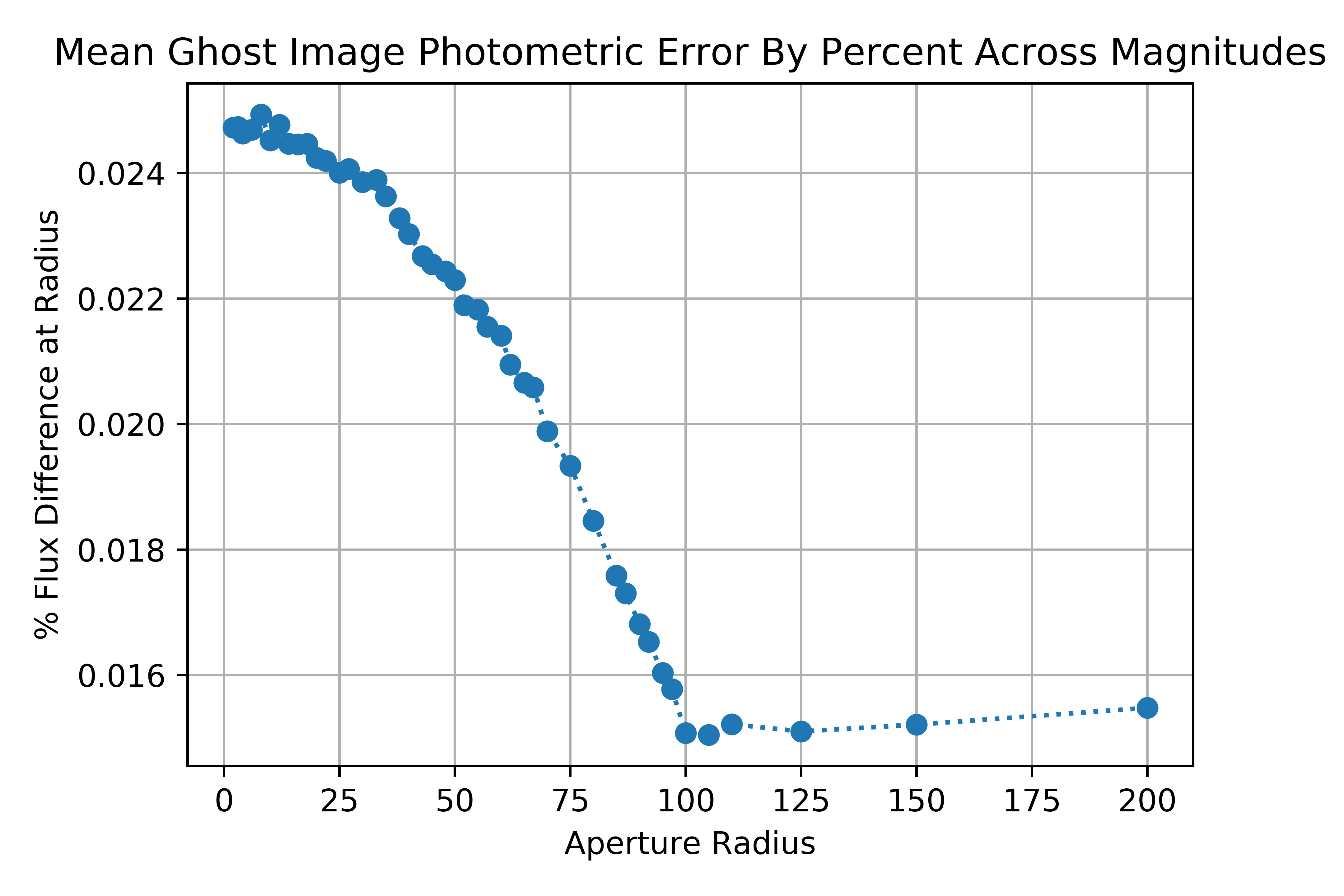}
    \includegraphics[scale=0.4]{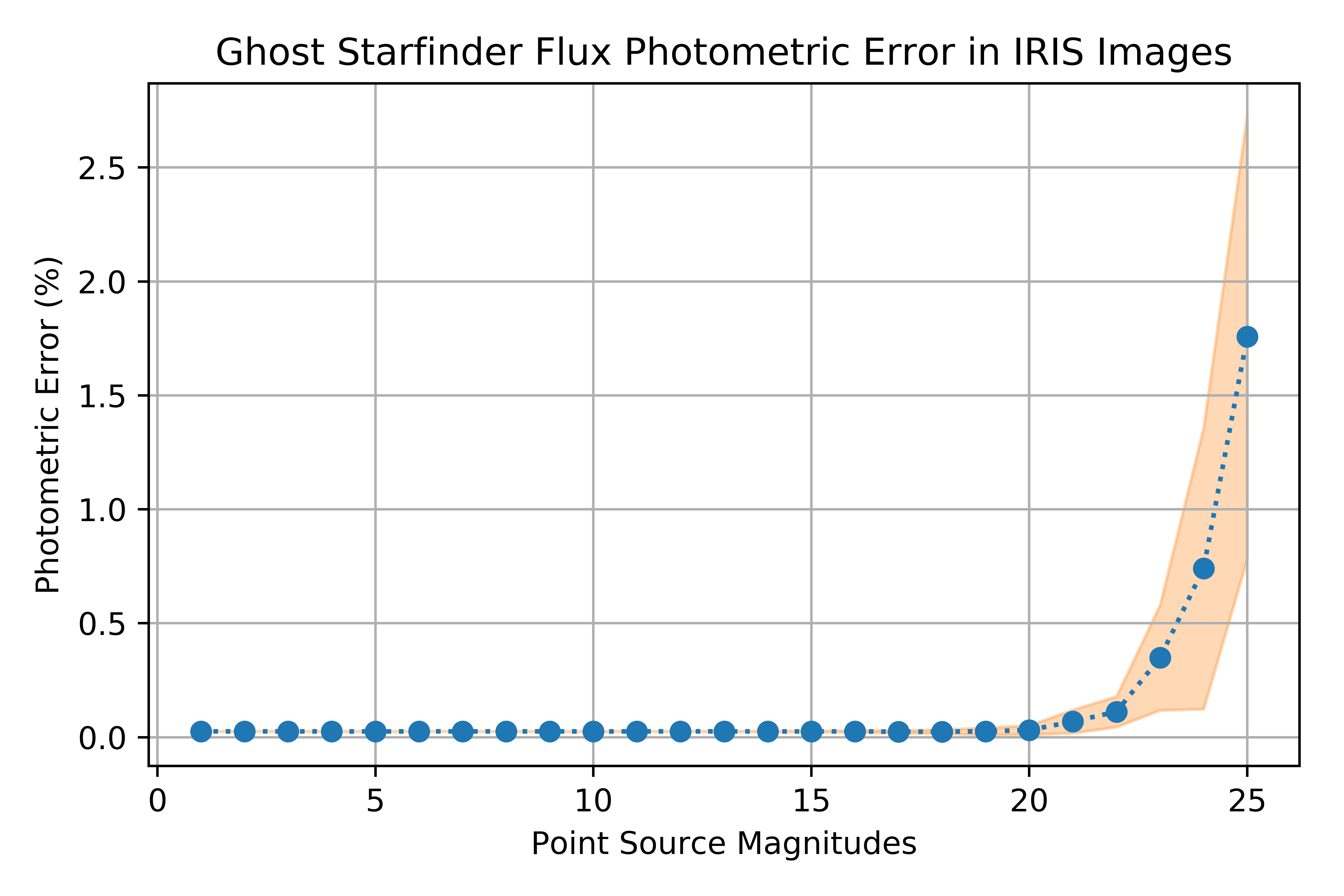}
    \caption{\textbf{Left:} Representation of simulated wedge ghost distance, d=0 pixels. \textbf{Middle:} The photometric error associated with the average flux of simulated point sources as a function of radius. In this case overall photometric error is decreased, as the wedge ghost is located on its source. \textbf{Right:} The photometric error associated with each magnitude point source simulation for high SNR as calculated by \textit{Starfinder}.}
    \label{fig:r0ghost}
\end{figure}

We simulate the photometric impact of the wedge filter ghost image assuming no filter-wedge angle, resulting in a distance of 0 pixels. Figure \ref{fig:r0ghost} illustrates the results of this simulation.

This scenario effectively eliminates the wedge filter ghost, as the flux from the wedge filter ghost is reintroduced to its source flux. This means the dominating ghost over the source is the entrance window ghost, and the general shape of the aperture photometry plot reflects that of Figure 1 for this reason (but includes the ADC ghost flux redistribution in this case). Maximum photometric error from aperture photometry is recorded at 0.0249\%, with \textit{Starfinder} results yielding 0.0251$\pm$0.0004\% photometric error for this case.

Table \ref{table:2} summarizes the photometric impact of various wedge-ghost image configurations.

\begin{table}[H]
\centering

\begin{tabular}{c|c|c|c}
    \multicolumn{4}{ c }{Wedge Filter Ghost Location Simulation Results}\\
    \hline
     & Figure Depiction & \tabcell{Wedge Ghost \\ Distance (Pixels)} & Photometric Error    \\
     \hline
     Standard Case& \tabcell{\includegraphics[scale=0.15]{ExtendedEntranceGhostComparisons.png}} & 16 & 0.290$\pm$0.002\%  \\
     \hline
     Ghost at Source & \tabcell{\includegraphics[scale=0.15]{WedgeGhostR0.png}} & 0 & 0.0251$\pm$0.0004\%  \\
     \hline
     Near Ghost & \tabcell{\includegraphics[scale=0.15]{ExtendedWedgeR10Ghost.png}} & 10 & 0.361$\pm$0.001\%.  \\
     \hline
     Far Ghost & \tabcell{\includegraphics[scale=0.15]{FarWedgeGhost.png}} & 50 & 0.276$\pm$0.001\%  \\
    
\end{tabular}
\label{table:2}
\caption{Photometric error associated with each simulation varying the distance of the wedge filter ghost image.}
\end{table}

\end{document}